\documentclass[11pt,letter,floatfix]{article}
\pdfoutput=1
\usepackage{jcappub}
\usepackage{amssymb}
\usepackage{gensymb}
\usepackage{graphicx}
\usepackage{amsmath}
\usepackage{wasysym}
\usepackage{verbatim}

\usepackage{graphicx}
\usepackage{lineno}                   
\usepackage{amsmath}                  
\usepackage{dcolumn}                  
\usepackage{bm}                       
\usepackage{xspace}                   
\usepackage{hyperref}                 
\usepackage{gensymb}

\usepackage{subfigure}

\usepackage{epsfig}
\usepackage{hyperref}
\usepackage{comment}
\usepackage{color}
\usepackage{cleveref}
\usepackage{multirow}
\usepackage{capt-of}
\usepackage{booktabs,rotating}
\usepackage{graphicx}
\usepackage{gensymb}
\usepackage{longtable}
\usepackage{undertilde}
\usepackage{float}

\hypersetup{
  colorlinks=true,
  linkcolor=blue,
  citecolor=blue,
  urlcolor=blue,
}

\begin{document}
%
\title{A Search for Dark Matter in the Galactic Halo with HAWC}
\emailAdd{hgayala@psu.edu} 
\emailAdd{jpharding@lanl.gov}
\author[a]{A.U.~Abeysekara}

\author[b]{A.M.~Albert}

\author[c]{R.~Alfaro}

\author[d]{C.~Alvarez}

\author[d]{R.~Arceo}

\author[e]{J.C.~Arteaga-Vel\'{a}zquez}

\author[c]{D.~Avila Rojas}

\author[f]{H.A.~Ayala Solares}

\author[c]{A.~Becerril}

\author[c]{E.~Belmont-Moreno}

% ...

\affiliation[a]{Department of Physics and Astronomy, University of Utah, Salt Lake City, UT, USA }

\affiliation[b]{Physics Division, Los Alamos National Laboratory, Los Alamos, NM, USA }

\affiliation[c]{Instituto de F\'{i}sica, Universidad Nacional Aut\'{o}noma de M\'{e}xico, Ciudad de Mexico, Mexico }

\affiliation[d]{Universidad Aut\'{o}noma de Chiapas, Tuxtla Guti\'{e}rrez, Chiapas, M\'{e}xico}

\affiliation[e]{Universidad Michoacana de San Nicol\'{a}s de Hidalgo, Morelia, Mexico }

\affiliation[f]{Department of Physics, Pennsylvania State University, University Park, PA, USA }

% ...

\emailAdd{nrodd@mit.edu} 
\emailAdd{bsafdi@mit.edu} 
\affiliation[cc]{Center for Theoretical Physics, Massachusetts Institute of Technology, Cambridge, MA 02139} 
\affiliation[dd]{Leinweber Center for Theoretical Physics, University of Michigan, Ann Arbor, MI 48109}

\date{\today}

\abstract{The High Altitude Water Cherenkov (HAWC) gamma-ray observatory is a wide field-of-view observatory sensitive to 500 GeV \textendash\ 100 TeV gamma rays and cosmic rays. 
With its observations over 2/3 of the sky every day, the HAWC observatory is sensitive to a wide variety of astrophysical sources, including possible gamma rays from dark matter. Dark matter annihilation and decay in the Milky Way Galaxy should produce gamma-ray signals across many degrees on the sky. The HAWC instantaneous field-of-view of 2 sr enables observations of extended regions on the sky, such as those from dark matter in the Galactic halo. Here we show limits on the dark matter annihilation cross-section and decay lifetime from HAWC observations of the Galactic halo with 15 months of data. These are some of the most robust limits on TeV and PeV dark matter, largely insensitive to the dark matter morphology. These limits begin to constrain models in which PeV IceCube neutrinos are explained by dark matter which primarily decays into hadrons.}

\subheader{LA-UR-17-29899, LCTP-17-01, MIT-CTP 4951}

\maketitle

\def\simgt{\mathrel{\lower2.5pt\vbox{\lineskip=0pt\baselineskip=0pt
           \hbox{$>$}\hbox{$\sim$}}}}
\def\simlt{\mathrel{\lower2.5pt\vbox{\lineskip=0pt\baselineskip=0pt
           \hbox{$<$}\hbox{$\sim$}}}}
\makeatother

\newcommand{\be}{\begin{equation}}
\newcommand{\ee}{\end{equation}}
\newcommand{\bea}{\begin{eqnarray}}
\newcommand{\eea}{\end{eqnarray}}
\newcommand{\Fig}[1]{Fig.~\ref{#1}}
\newcommand{\Figs}[2]{Fig.~\ref{#1} and \ref{#2}}
\newcommand{\Eq}[1]{Eq.~(\ref{#1})}
\newcommand{\Eqs}[2]{Eqs.~(\ref{#1}) and (\ref{#2})}
\newcommand{\Sec}[1]{Sec.~\ref{#1}}
\newcommand{\Secs}[2]{Secs.~\ref{#1} and \ref{#2}}
\newcommand{\App}[1]{App.~\ref{#1}}
\newcommand{\vev}[1]{\langle #1 \rangle}

\newcommand{\aaps}{{Astron.~Astrophys.~Supp.}}
\newcommand{\araa}{{Annu.~Rev.~Astron.~Astrophys.}}
\newcommand{\aap}{{Astron.~Astrophys.}}
\newcommand{\apjl}{{Astrophys.~J.~Lett.}}
\newcommand{\aj}{{Astron.~J.}}
\newcommand{\mnras}{{Mon.~Not.~R.~Astron.~Soc.}}
\newcommand{\apjs}{Astrophys. J. Supplement}
\newcommand{\pasa}{PASA}

\newcommand{\B}{{\cal B}}

\newcommand{\mPl}{m_{\rm Pl}}
\newcommand{\LL}{\mathcal{L}}
\newcommand{\OO}{\mathcal{O}}
\newcommand{\OOMSSM}{\mathcal{O}_{\rm MSSM}}
\newcommand{\BR}{\textrm{BR}}
\newcommand{\sslash}[1]{\ensuremath\raisebox{-0.00cm}{{\small\slash}}\hspace{-0.21cm}#1\/}
\newcommand{\dd}[1]{\frac{\partial}{\partial #1}}

\newcommand\nr[1]{{\bf {\color{red}(NR: #1)}}}

\newcommand{\draftnote}[1]{\textbf{#1}}

\section{Introduction}
\subsection{Heavy Dark Matter}
For all the evidence pointing toward the existence of particle dark matter (DM), no observations have yet indicated a preferred mass scale in the dark sector.
The prevailing view is that DM should be associated with new physics at the electroweak scale; however, this has so far not been realized experimentally. 
This motivates a re-examination of the different mass scales where DM could exist.
%
%

If the DM in the Universe is composed of a particle with a mass well above the electroweak scale (\textgreater 1 TeV), then its non-gravitational nature may only be discoverable through indirect detection.
At these masses direct production of DM at a collider becomes inaccessible. Furthermore the expected direct detection rate in many models becomes negligible due to the low DM number density and small DM-nucleon cross-sections.
Yet if the DM can decay or annihilate to standard model (SM) final states, the stable cosmic rays resulting from these processes can leave detectable imprints in the sky.
HAWC, with its ability to search for TeV gamma-rays over large portions of the sky, is ideally placed to search for such signals of heavy DM (mass \textgreater 1 TeV), and in the absence of a signal set strong limits on the DM parameter space.
This is the strategy we pursue to set limits on heavy DM in this work.

Theoretically, heavy DM is also well motivated.
As a concrete example, if the dark sector consists only of a confining gauge theory, then there will be a dark glueball in the spectrum with a mass at the scale of confinement, which can be well above the electroweak scale
%
%
~\cite{Faraggi:2000pv,Boddy:2014yra,Forestell:2016qhc,Halverson:2016nfq,Cohen:2016uyg,Acharya:2017szw,Soni:2017nlm}.

At these masses it is common to focus on decaying DM, as the simplest models where the DM annihilates cannot produce the observed DM thermal relic abundance without being in violation of the unitarity bound~\cite{Griest:1989wd,Hui:2001wy}, see also~\cite{Beacom:2006tt}.
Nevertheless, models exist which evade these bounds, see e.g.~\cite{Berlin:2016gtr,Berlin:2016vnh}, and so limits on DM annihilation at these masses should still be considered.

A further impetus in searching for DM signals at these masses comes from the detection of an astrophysical high-energy neutrino flux at IceCube~\cite{Aartsen:2013bka,Aartsen:2013jdh,Aartsen:2015knd,Aartsen:2015rwa}.
At present no point sources have been detected in the data~\cite{Aartsen:2013uuv,Aartsen:2014cva,Aartsen:2016oji} and whilst a number of purely astrophysical explanations of the flux remain, see e.g.~\cite{Hooper:2016jls} and in particular~\cite{Murase:2016gly} for a recent summary, the presence of a large contribution from DM cannot at present be excluded using the IceCube data alone.
Accordingly, a number of models have been developed which explain the high-energy neutrino flux with DM~\cite{Esmaili:2013gha,Feldstein:2013kka,Ema:2013nda,Zavala:2014dla,Bhattacharya:2014vwa,Higaki:2014dwa,Rott:2014kfa,Fong:2014bsa,Dudas:2014bca,Ema:2014ufa,Murase:2015gea,Anchordoqui:2015lqa,Boucenna:2015tra,Ko:2015nma,Aisati:2015ova,Kistler:2015oae,Chianese:2016opp,Fiorentin:2016avj,Dev:2016qbd,DiBari:2016guw,Kalashev:2016cre,Chianese:2016smc,Cohen:2016uyg}.
Generally these models seek to explain the highest energy neutrinos observed by IceCube---the so-called High Energy Starting Events (HESE)---however some also deal with a putative excess in the lower energy Medium Energy Starting Events (MESE) \cite{Aartsen:2014muf}, which has been discussed in the context of DM~\cite{Chianese:2016kpu} and astrophysics~\cite{Murase:2015xka,Palladino:2016xsy}.
Nevertheless, in these models the DM decays will generically also produce photons which can be searched for in other experiments.
In particular~\cite{Cohen:2016uyg} considered the limits obtained with the {\it Fermi}-LAT, and demonstrated that models which seek to explain the majority of the IceCube flux appear to be in tension with the gamma-ray data.
HAWC, with its ability to study gamma-rays at even higher energy than the {\it Fermi}-LAT, can also constrain this parameter space, as we demonstrate in this work.

\subsection{Using HAWC to Constrain DM}
The Galactic center is expected to have the largest DM signal in the sky. However, the Galactic center transits at the edge of the HAWC field-of-view, so we instead consider a region near the Galactic center but extended to higher declinations where HAWC observations are more favorable. In particular, in the work, we consider the region of the Northern Fermi Bubble.  The region of the Northern Fermi Bubble is particularly useful in constraining DM annihilations and decays from the Galactic halo for HAWC because:
\begin{itemize}
\item It is a region which extends outward from the Galactic center;
\item It has large solid angle for integrating the DM gamma-ray flux;
\item It extends to declinations of 8 degrees North (culminating at 11 degrees zenith for HAWC), so the analysis is not strongly dependent on the HAWC sensitivity at declinations with low sensitivity;
\item The systematic errors are well-characterized, including an understanding of the background exclusion mask and well-behaved background significance distribution~\cite{Abeysekara:2017wzt}; and
\item It is known that HAWC sees no significant gamma-ray excess in this region, which minimizes confusion between DM and standard astrophysical sources.
\end{itemize}
The DM limits do not strongly depend on the chosen region-of-interest, but the characterized systematics and lack of astrophysical gamma rays in the Northern Fermi Bubble region-of-interest makes the analysis robust. Additionally, this is an excellent example of how DM limits are based on flux limits, so a calculated flux limit can be used to limit DM annihilation and decay.

In this work, we discuss the region of interest for the analysis as well as the methods used to calculate the background. We use these values to calculate a flux limit on emission from a region toward the Galactic center. We then derive a gamma-ray flux from DM annihilation or decay which could be searched for in this region. Finally, we present the lack of observed signal as a limit on the DM annihilation cross-section and decay lifetime and discuss the implications of these limits in the context of other astrophysical experiments.

\section{Methods}\label{methods}
\subsection{Procedure}
The present work searches for a gamma-ray signal from the direction of the Galactic center (GC) using data from the HAWC observatory. A  search for gamma-ray signals from the  Northern Fermi Bubble region has been presented in Ref.~\citep{Abeysekara:2017wzt}. The same analysis procedure, data, analysis bins, and search region are used here. 
The data for the analysis corresponds to the dates between 2014 November 27th and 2016 February 11th.
Since it is required that the detector is stable for a period of 24 hours for this analysis, we use only the 290 days of the data over the 15 month period which meet this criterion.
The data are binned according to $fHit$,the fraction of functioning photomultipliers triggered in an air shower event.  Table 1 in ~\cite{Abeysekara:2017wzt} defines the values for each analysis bin. For details about the detector see Ref.~\cite{HAWCCrab}.

The main challenge of the analysis, as discussed in Ref.~\cite{Abeysekara:2017wzt}, is to estimate the background. There are three main factors that need to be considered for this: to distinguish the air showers produced by hadronic cosmic rays from those produced by gamma rays; to find the isotropic flux of hadronic cosmic rays through the direct integration (DI) method \citep{Atkins:2003}; and to remove the effects of the large-scale cosmic-ray anisotropy \citep{Abeysekara:2014}.

\begin{figure}
\centering
\includegraphics[width=0.8\textwidth]{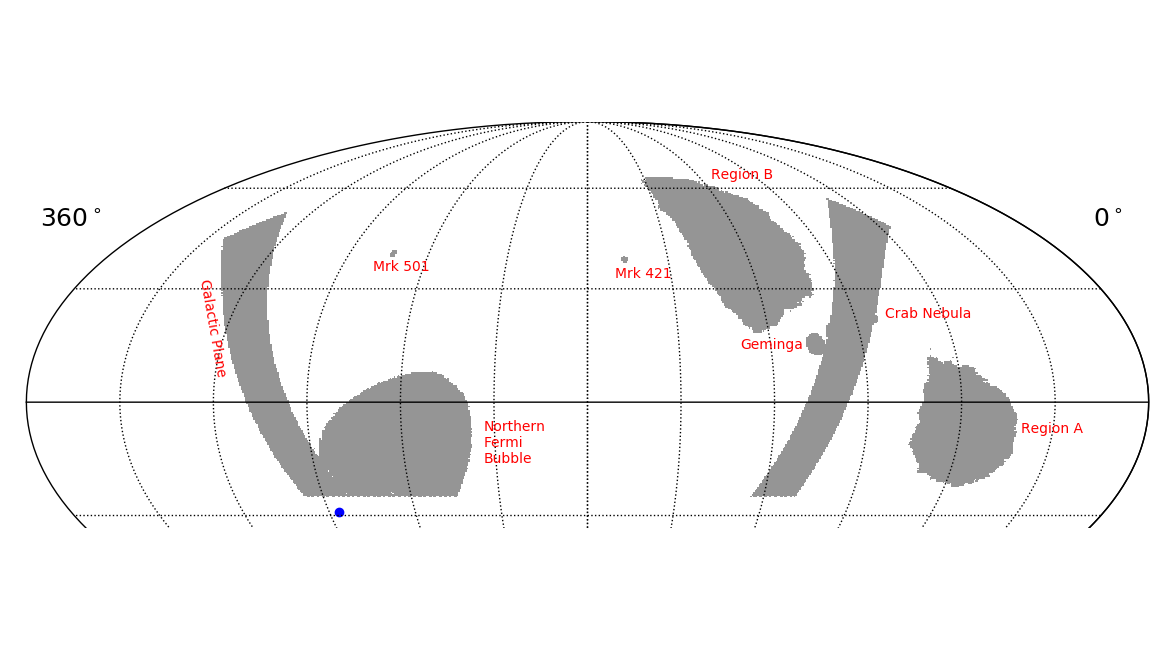}
\caption{Regions masked when calculating the isotropic background, plotted in equatorial coordinates. The masked regions include the Galactic plane, which contains many sources of gamma-ray emission, and the bright HAWC sources Mrk 421, Mrk 501, Geminga, and the Crab Nebula. The cosmic-ray excesses in the anisotropy Regions A and B~\cite{Abeysekara:2014} are masked out to give consistent background calculation. Also, the signal region of the Northern Fermi Bubble is masked out so that any possible signal does not contaminate the background. The Galactic center is marked with a blue point}
\label{fig:ROI}
\end{figure}

For the first factor, gamma-hadron separation cuts are applied to the data. These separators select the gamma-like showers~\cite{HAWCCrab}. For more details on the values of these cuts, see Table 2 of Ref.~\cite{Abeysekara:2017wzt}.
For the isotropic flux estimation, a 24 hour integration period is used for the DI method. This is to ensure that the analysis is sensitive to large structures.  
A masking region around known sources is used in order to avoid contamination from the signal from these sources into the estimation of the isotropic background. Figure \ref{fig:ROI} shows the region used for the masking.

\begin{figure}
\centering
\includegraphics[width=0.8\textwidth]{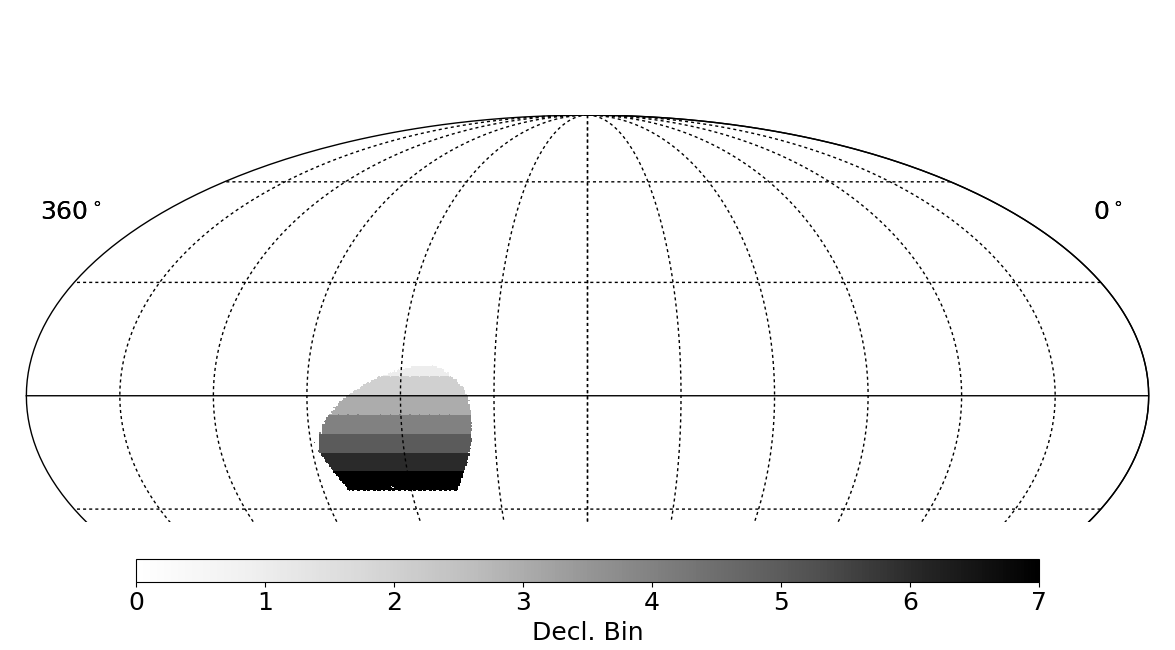}
\caption{Norther Fermi Bubble divided into declination bands, plotted in equatorial coordinates. These are the signal regions used in this analysis.}
\label{fig:decBands}
\end{figure}

A subtraction procedure is implemented to mitigate the effects of large-scale anisotropy in the data. A more detailed discussion of these analyses may be found in Ref.~\cite{Abeysekara:2017wzt}.
 
\subsection{HAWC Differential Flux Upper Limits near the Galactic Center Region}

Instead of adding the total flux inside the search region as in Ref.~\cite{Abeysekara:2017wzt}, the search region is divided in six declination bands of $5\degree$ each, as shown in Figure \ref{fig:decBands}, and a total possible DM flux in each declination band is calculated. This division is done in order to take into account the expected spatial dependence of a DM signal and the declination dependence of the sensitivity of the HAWC detector.

\begin{figure}
\centering
\includegraphics[width=0.8\textwidth]{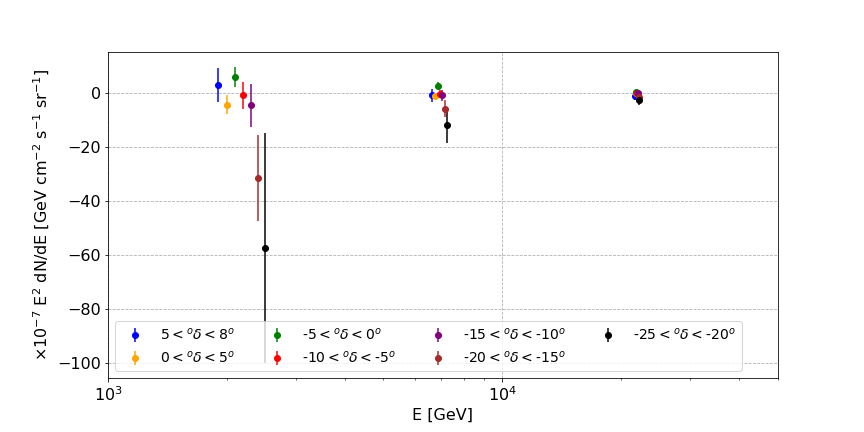}
\caption{Gamma-ray flux in each declination band. These values are given in Table~\ref{tab:uplims}.}
\label{fig:fluxes}
\end{figure}

\begin{figure}
\centering
\includegraphics[width=0.8\textwidth]{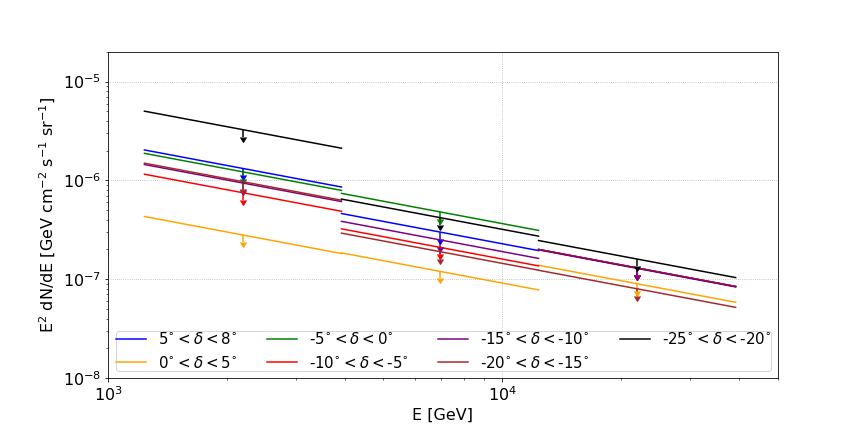}
\caption{Gamma-ray flux upper limit in each declination band. These values are given in Table~\ref{tab:uplims}.}
\label{fig:uplims}
\end{figure}

The number of excess events in each declination band is shown in Table \ref{tab:excess}. Note that the distribution of significances is consistent with Gaussian background distribution.
The fluxes in each declination band are calculated using the same procedure as in Ref.~\cite{Abeysekara:2017wzt}.  
Figures \ref{fig:fluxes}, \ref{fig:uplims}, and Table \ref{tab:uplims} show the results of this analysis.
The energy bins have a size of $\Delta \log (E/1\text{\,TeV}) = 0.5$. This size corresponds to the approximate width of the energy response histograms of the detector (See Figure 8 of \cite{Abeysekara:2017wzt}).  The fluxes in each energy bin are calculated as a weighted average of the overlapping analysis bins. The first energy bin is centered at 2.2\,TeV, which is the median energy of analysis bin $f_1$. Although the analysis of Ref.~\cite{Abeysekara:2017wzt} included an overflow bin for all energies above 39 TeV, that is not useful in constraining spectral information and is not included here.
Note that the highest-declination search bin only goes from 5 to 8 degrees declination because the Northern Fermi Bubble cuts off at higher declinations.

%

The process of DI background subtraction used in HAWC removes all isotropic backgrounds, primarily hadrons but also any isotropic gamma-ray signal. For a spatially-dependent gamma-ray signal spread over a large region of the sky, such as a signal from the Milky Way DM halo, the portion of the gamma-ray signal that is outside the search region will be subtracted as well. Because of this, we account for only the difference between the amount of signal expected within the search region to that expected outside in our analysis. However, because the DM signal considered in this analysis peaks toward the GC, this residual flux is large enough to set strong constraints.

\begin{sidewaystable}[tph!]
\begin{center}
\begin{tabular}{|c|c|c|c|c|c|c|c|}
\hline
Analysis & \multicolumn{7}{c|}{Excess in Declination Band (Counts in 290 Days)} \\ \cline{2-8}
fHit bin & 5\degree$\ -\ $8\degree & 0\degree$\ -\ $5\degree  &-5\degree$\ -\ $0\degree & -10\degree$\ -\ $-5\degree & -15\degree$\ -\ $-10\degree & -20\degree$\ -\ $-15\degree  & -25\degree$\ -\ $-20\degree  \\
\hline
1 & 1200$\pm$1500  & -900$\pm$ 2760 & -57$\pm$ 2835 &-1300$\pm$2600  & -390$\pm$2160 & -2300$\pm$1600 & -390$\pm$1080  \\
2 & 56$\pm$510 & -1630$\pm$910 & 2000 $\pm$900 & 690$\pm$860 & -710$\pm$710 & -790$\pm$530 & -690$\pm$350\\
3 &  -93$\pm$235 & 55$\pm$421 & 840$\pm$430& -310$\pm$400& 160$\pm$330 &16$\pm$240 & -87$\pm$153\\
4 &  -97$\pm$99 & -17$\pm$178 & -170$\pm$180& -7.6$\pm$166.7 & 20$\pm$130 & -83$\pm$96 & -78$\pm$60\\
5 &  40$\pm$76 & 23 $\pm$137 & 220$\pm$140& -120$\pm$130& -77$\pm$100 & -25$\pm$73 & 60$\pm$40\\
6 &  -52$\pm$39 & 6.2$\pm$70.0 & -56$\pm$72& 110$\pm$70 & 42$\pm$51 & 1.0$\pm$35.9 & 4.6$\pm$21.1 \\
7 &  4.0$\pm$47.7 & 74$\pm$86 & -180$\pm$90 & -15$\pm$81 & 36$\pm$63 & 6.2$\pm$42.5 & 33$\pm$25\\
 \hline
\end{tabular}
\end{center}
\caption{Number of excess events in each declination and analysis bin. See Table 1 of Ref.~\cite{Abeysekara:2017wzt} for detailed description of the analysis bins.}
\label{tab:excess}
\bigskip\bigskip

\begin{center}
\begin{tabular}{|c|c|c|c|c|c|c|c|}
\hline
Energy & \multicolumn{7}{c|}{Fluxes and Upper limits per Declination Band $\left(E^2d^2F/dEd\Omega\right) \left(\times 10^{-7} \text{GeV}\,\text{cm}^{-2}\,\text{s}^{-1}\,\text{sr}^{-1}\right)$} \\\cline{2-8}
 Range (TeV) & 5\degree$\ -\ $8\degree & 0\degree$\ -\ $5\degree  &-5\degree$\ -\ $0\degree & -10\degree$\ -\ $-5\degree & -15\degree$\ -\ $-10\degree & -20\degree$\ -\ $-15\degree  & -25\degree$\ -\ $-20\degree  \\
\hline
1.2 - 3.9 & 2.8$\pm$6.3 (13.2) & -4.3$\pm$3.4 (2.8) & 5.9$\pm$3.8 (12.2) & -0.8$\pm$5.0 (7.5) & -4.5$\pm$8.0 (9.4)& -31.0$\pm$16.0 (9.7) & -57.0$\pm$42.0 (32.6) \\
3.9 - 12.4 & -0.9$\pm$2.3 (3.0) & -1.1$\pm$1.2 (1.2)& 2.8$\pm$1.2 (4.8) & -0.4$\pm$1.5 (2.1) & -0.9$\pm$2.0 (2.5) & -5.9$\pm$3.1 (1.9) & -12.0$\pm$6.6 (4.2) \\
12.4 - 39.1 & -1.2$\pm$1.3 (1.3) &  -0.1$\pm$0.6 (0.9) & 0.3$\pm$0.6 (1.3) & 0.1$\pm$0.7 (1.3) & 0.0$\pm$0.8 (1.3) & -1.6$\pm$1.1(0.8) & -2.6$\pm$2.0 (1.6) \\
\hline
\end{tabular}
\end{center}
\caption{Fluxes and 95$\%$ confidence level flux upper limits calculated in each declination band. Upper limits are inside the parenthesis.}
\label{tab:uplims}

\end{sidewaystable}
\clearpage

\section{Gamma-Ray Flux from Dark Matter Near the Galactic Center}
\subsection{Dark Matter Annihilation and Decay}
To calculate the gamma-ray flux from DM annihilation or decay, information both about the particle nature of DM and its astrophysical distribution must be known. The gamma-ray flux is proportional to the square of the DM density for annihilation or the DM density itself for decay, integrated over the line-of-sight through the DM halo from the observer. This integral (with appropriate normalizations) is referred to as the annihilation $J$-factor or the decay $D$-factor. These are given by:
\begin{eqnarray}
J&=&\int_{\Delta\Omega}d\,\Omega\int d\,x\ \rho^2(r_{\rm{gal}}(\theta,x))\enspace,
\\
D&=&\int_{\Delta\Omega}d\,\Omega\int d\,x\ \rho(r_{\rm{gal}}(\theta,x))\enspace.
\end{eqnarray}
The distance from the center of the galaxy is given by
\begin{equation}
r_{\rm{gal}}(\theta,x)=\sqrt{R_{\odot}^2-2xR_{\odot}\cos(\theta)+x^2}
\end{equation}
at a polar angle $\theta$ between the observer line-of-sight and the GC. The solid angle of the observation is given by $\Delta\Omega$. In this paper, we use the distance from the GC to the Earth as $R_{\odot}=8.2\rm\,kpc$ and a DM density at the Earth of $\rho_{\odot}=0.39\rm\,GeV cm^{-3}$ as in Ref.~\cite{Catena:2009mf}.

We consider a range of DM masses $M_\chi$ above 7 TeV for annihilating DM and above 14 TeV for decaying DM. We also consider the spectrum of gamma rays produced in a single DM decay or annihilation $dN/dE$ for several DM annihilation or decay channels. The DM velocity-weighted cross-section for annihilation $\langle\sigma v\rangle$ or lifetime for DM decay $\tau$ are also needed. With these, the gamma-ray flux from DM annihilation or decay is given by
\begin{eqnarray}
\frac{dF_{\rm ann}}{dE d\Omega}&=&\frac{\langle\sigma v\rangle}{8\pi M_\chi^2}\frac{J}{\Delta\Omega}\frac{dN}{dE}\label{fann}\enspace,\\
\frac{dF_{\rm dec}}{dE d\Omega}&=&\frac{1}{4\pi M_\chi \tau}\frac{D}{\Delta\Omega}\frac{dN}{dE}\label{fdec}\enspace.
\end{eqnarray}

\subsection{Dark Matter Halo Profiles}
\begin{figure}
\centering
\includegraphics[width=0.7\textwidth]{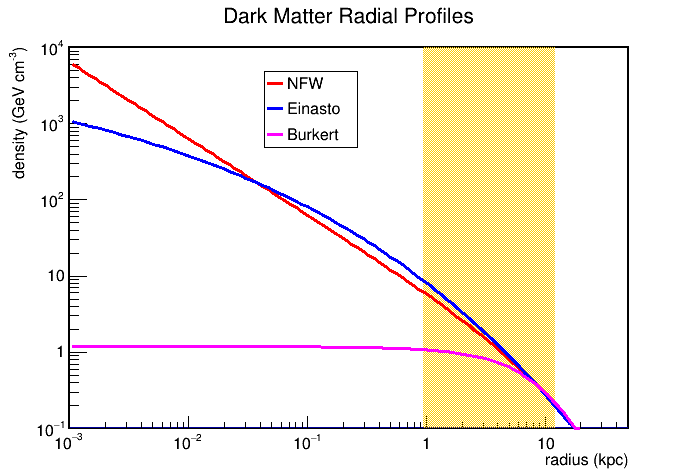}
\caption{The three DM profiles considered for the Milky Way. The red curve is the NFW profile, the blue curve is the Einasto profile, and the magenta curve is the Burkert profile. Note that the Burkert profile has a much more cored dark matter profile and does not reach as high a density at the GC that the cusped NFW and Einasto profiles do. However, the orange region shows the range of galactocentric radii considered in this analysis. Our limits here are primarily sensitive at larger radii, at a few kpc, where the three profiles have more similar densities. Therefore, our limits are not strongly dependent on the choice of DM profile.}
\label{DMprofileFig}
\end{figure}

While the presence of DM in the galaxy is well established, the precise density distribution of the DM is not. Therefore, we consider three different DM density profiles in this paper, including profiles peaked toward the GC and those with a density core toward the GC. N-body simulations of DM without baryons are typically fit with a Navarro-Frenk-White (NFW) DM profile~\cite{Navarro:1996gj} with distribution
\begin{equation}
\rho(r)=\frac{\rho_{s}}{(r/r_{s})(1+r/r_{s})^{2}}\enspace.
\end{equation}
For the Milky Way, the scale radius is on the order of $r_s=20\rm\,kpc$~\cite{Nesti:2013uwa}. Normalizing to the local DM density, this gives $\rho_s=0.318\rm\,GeVcm^{-3}$. A different fit to N-body simulations with a shallower core is given by the Einasto profile~\cite{Stadel:2008pn,Navarro:2008kc}. For the Milky Way, this profile is given by
\begin{equation}
\rho(r)=\rho_s
\exp\left[-\frac{2}{\alpha}\left(\left(\frac{r}{r_s}\right)^{\alpha}
    -1\right)\right]\enspace,
\end{equation}
with $\alpha=0.17$, $r_s=20\rm\,kpc$, and $\rho_s=0.0746\rm\,GeVcm^{-3}$~\cite{Pieri:2009je}. Observationally, the DM may also have a more cored profile near the GC, given by a Burkert profile~\cite{Burkert:1995yz}. This can be parameterized as
\begin{equation}
\rho(r)=\frac{\rho_{s}}{(1+r/r_{s})(1+(r/r_{s})^{2})}\enspace.
\end{equation}
For the Milky Way, $r_s=10\rm\,kpc$ and $\rho_s=1.187\rm\,GeVcm^{-3}$ for the Burkert profile~\cite{Nesti:2013uwa}.
A comparison of these profiles, including the peaked and cored features, can be seen in Figure~\ref{DMprofileFig}.
The $J$- and $D$-factors for each of these profiles in our signal region are tabulated in Table~\ref{jdtable}. 

\begin{figure}
\centering
\includegraphics[width=0.5\textwidth]{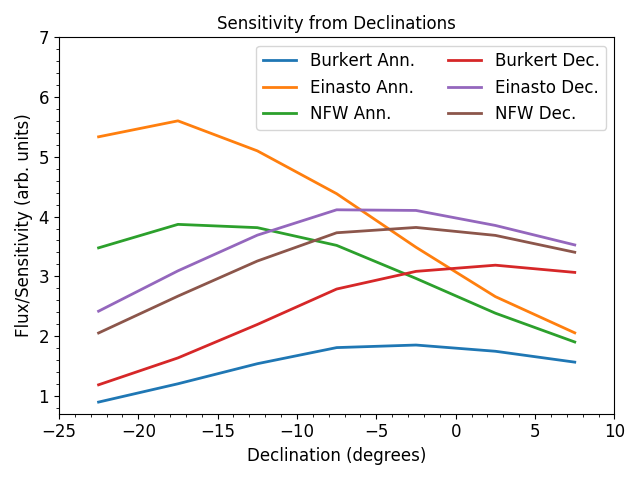}
\caption{The relative sensitivity of this analysis versus declination. On the y-axis, the DM flux in each declination is normalized by the HAWC sensitivity for that declination (from Ref.~\cite{Abeysekara:2017hyn}). This shows the contribution to the limit from each declination. For the Burkert annihilation profile and all three decay profiles, the peak contribution to the limit stems from declinations $>20\degree$ north of the Galactic center (declination $-29\degree$) that transit closer to the HAWC zenith.}
\label{DMSensFig}
\end{figure}

Although the three DM profiles considered here differ significantly near the GC, HAWC analysis is most sensitive to regions away from the GC where the three profiles are in general agreement. Therefore, this analysis is not as sensitive to profile shape as other DM annihilation analyses with the GC. DM decay analyses are also less sensitive to the shape of the DM profile, since they depend linearly on the DM density rather than quadratically. Figure~\ref{DMSensFig} demonstrates how much each declination band contributes to the limit for each DM profile. For Burkert annihilation profile and all the decay profiles, the peak sensitivity to DM for HAWC is over $20\degree$ from the GC.

\subsection{The Background-subtracted $J$- and $D$-factors}
The background calculation for the HAWC analysis uses a 24-hour DI technique to determine the cosmic-ray background~\cite{Abeysekara:2017wzt}. For an extended signal, like the DM one we consider here, there is a nontrivial amount of gamma-ray emission in the background region as well. Therefore, the gamma-ray excess limits calculated in section~\ref{methods} are actually a limit on the gamma-ray excess in the target region minus the gamma-ray excess in the background region (that is, the region outside of the mask). For each declination bin, therefore, we calculate the $J$-factor and $D$-factor over both the source and the background region. Our limit is calculated based on the difference of these two values (i.e. $J_{\rm src}-J_{\rm bkg}$ and $D_{\rm src}-D_{\rm bkg}$.) The values of these $J$- and $D$-factors are given in Table~\ref{jdtable}.
\begin{table}
\begin{center}
\begin{tabular}[t]{|c|r@{\ }c@{\ }l|c|c|c|c|c|c|}
  \hline
  \multirow{2}{*}{DM profile} & \multicolumn{3}{c|}{\multirow{2}{*}{Dec range}} & \multirow{2}{*}{$\frac{J_{\rm src}}{\Delta\Omega}$} & \multirow{2}{*}{$\frac{J_{\rm bkg}}{\Delta\Omega}$} & \multirow{2}{*}{$\frac{J_{\rm src}-J_{\rm bkg}}{\Delta\Omega}$} & \multirow{2}{*}{$\frac{D_{\rm src}}{\Delta\Omega}$} & \multirow{2}{*}{$\frac{D_{\rm bkg}}{\Delta\Omega}$} & \multirow{2}{*}{$\frac{D_{\rm src}-D_{\rm bkg}}{\Delta\Omega}$} \\
&\multicolumn{3}{c|}{}&&&&&&\\
  \hline
Burkert & 5\degree & $-$ &  8\degree & 10.3 & 5.2 & 5.1 & 27.9 & 17.9 & 10.0 \\
 & 0\degree & $-$ &  5\degree & 11.8 & 5.5 & 6.3 & 29.8 & 18.3 & 11.5 \\
 & -5\degree & $-$ &  0\degree & 13.9 & 6.1 & 7.8 & 32.1 & 19.1 & 13.0 \\
 & -10\degree & $-$ &  -5\degree & 16.1 & 6.7 & 9.4 & 34.4 & 19.9 & 14.5 \\
 & -15\degree & $-$ &  -10\degree & 18.2 & 7.3 & 11.0 & 36.4 & 20.7 & 15.7 \\
 & -20\degree & $-$ &  -15\degree & 20.1 & 7.9 & 12.2 & 38.1 & 21.5 & 16.6 \\
 & -25\degree & $-$ &  -20\degree & 21.5 & 8.5 & 13.0 & 39.4 & 22.2 & 17.2 \\
\hline
Einasto & 5\degree & $-$ &  8\degree & 12.6 & 6.0 & 6.7 & 30.2 & 18.7 & 11.5 \\
 & 0\degree & $-$ &  5\degree & 16.2 & 6.7 & 9.6 & 33.4 & 19.4 & 13.9 \\
 & -5\degree & $-$ &  0\degree & 22.8 & 8.1 & 14.7 & 38.0 & 20.7 & 17.3 \\
 & -10\degree & $-$ &  -5\degree & 33.0 & 10.2 & 22.8 & 43.5 & 22.1 & 21.4 \\
 & -15\degree & $-$ &  -10\degree & 49.9 & 13.4 & 36.5 & 50.2 & 23.8 & 26.4 \\
 & -20\degree & $-$ &  -15\degree & 75.5 & 18.6 & 56.9 & 57.2 & 25.8 & 31.4 \\
 & -25\degree & $-$ &  -20\degree & 106.2 & 28.7 & 77.5 & 63.4 & 28.3 & 35.1 \\
\hline
NFW & 5\degree & $-$ &  8\degree & 12.0 & 5.8 & 6.2 & 30.0 & 18.9 & 11.1 \\
 & 0\degree & $-$ &  5\degree & 14.9 & 6.3 & 8.6 & 32.8 & 19.5 & 13.3 \\
 & -5\degree & $-$ &  0\degree & 20.0 & 7.5 & 12.5 & 36.8 & 20.6 & 16.1 \\
 & -10\degree & $-$ &  -5\degree & 27.4 & 9.1 & 18.3 & 41.4 & 22.0 & 19.4 \\
 & -15\degree & $-$ &  -10\degree & 38.6 & 11.3 & 27.3 & 46.7 & 23.4 & 23.3 \\
 & -20\degree & $-$ &  -15\degree & 54.0 & 14.7 & 39.3 & 52.1 & 25.0 & 27.1 \\
 & -25\degree & $-$ &  -20\degree & 70.9 & 20.4 & 50.5 & 56.7 & 26.8 & 29.8 \\
  \hline
\end{tabular}
\caption[Table of the $J$- and $D$-factors.]{
Table of the $J$- and $D$-factors in each bin of this analysis. The factors are shown for the Burkert, Einasto, and NFW profiles in each declination bin. The constrained values of $(J_{\rm src}-J_{\rm bkg})/\Delta\Omega$ and $(D_{\rm src}-D_{\rm bkg})/\Delta\Omega$ are given as well. ($J/\Delta\Omega$) are given in units of {$10^{21}\rm GeV^2cm^{-5}$} and ($D/\Delta\Omega$) are given in units of {$10^{21}\rm GeV cm^{-2}$}.
\label{jdtable}}
\end{center}
\end{table}
HAWC is currently working on background calculation methods which would minimize the amount of gamma-ray flux in the chosen background region, so the limits should correspondingly improve when those methods have matured.

\subsection{Dark Matter Spectra}\label{spectra}

When discussing DM spectra, it is typical to assume a dominant annihilation or decay channel (e.g. the DM annihilates into a pair of $Z$ bosons 100 percent of the time). The initial particles created in the channel are then allowed to decay into stable particles (including neutrinos, electrons, positrons, protons, and photons). 
There are three main sources of gamma rays from heavy DM annihilation or decays:
\begin{enumerate}
\item ``Prompt'' photons originating from the initial interaction occurring within the Milky Way halo. These include all primary photons produced in the initial annihilation or decay and the photons produced in the decays of the initial products.
\item Secondary photons up-scattered from the cosmic microwave background or interstellar radiation fields due to inverse-Compton scattering off the prompt electrons and positrons
\item Secondary photons produced from the cascades that develop due to interactions of the primary photons with background photons (e.g. the cosmic microwave background). This effect is only important over distance scales greater than $\mathcal{O}({\rm Mpc})$, much further than the Galactic center. 
\end{enumerate}
All three of these contribute to setting a limit in the 200 MeV to 2 TeV energy range of the {\it Fermi}-LAT, as shown in Ref.~\cite{Cohen:2016uyg}.
We follow the prescription in that reference for the various contributions, using the {\sc PYTHIA} 8.219 software~\cite{Sjostrand:2006za,Sjostrand:2007gs,Sjostrand:2014zea} with electroweak corrections enabled~\cite{Christiansen:2014kba}. However, here we will only consider the prompt emission as at TeV energies it is by far the largest component at these distances and energies.
Importantly, neglecting the contributions from secondary emission processes is conservative as it decreases our total expected flux and also reduces our sensitivity to the systematics associated with calculating these secondaries.
In addition we neglect the attenuation of these prompt gamma rays over Galactic scales.
At the energies considered and within our region of interest the attenuation is expected to be a small effect (\textless 5\%)~\cite{Esmaili:2015xpa}.

\section{Results}

 \subsection{Combining Limits in Declination and Energy}
We use the $J$- and $D$-factors for each morphological assumption along with the spectra described in section~\ref{spectra} to calculate the expected gamma-ray flux from DM via equations \ref{fann} and \ref{fdec}. These values are calculated in each declination bin for each energy bin. With the spectrum and DM mass fixed, we have only the cross-section or lifetime as a global free scaling parameter on the dark matter flux in each bin. We perform a $\chi^2$ test comparing these dark matter fluxes with the measured gamma-ray flux values for each declination/energy bin (from Table~\ref{tab:uplims}), following Appendix B of reference~\cite{Albert:2017vtb}. This gives us the 95\% CL limit on either the dark matter lifetime or cross-section for that mass and channel. We then repeat this over several masses and channels to get the results shown below.

\subsection{Dark Matter Limits}

\begin{figure*}[t]
 \subfigure{
    \hspace{-0.3cm}\includegraphics[width=7.9cm]{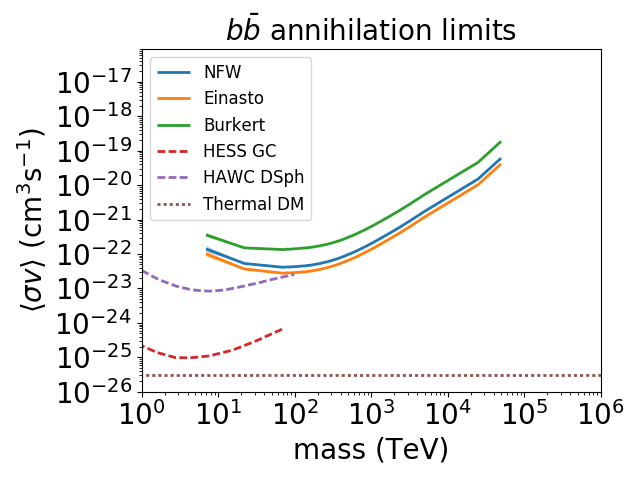}
  }
  \subfigure{
    \hspace{-0.5cm}\includegraphics[width=7.9cm]{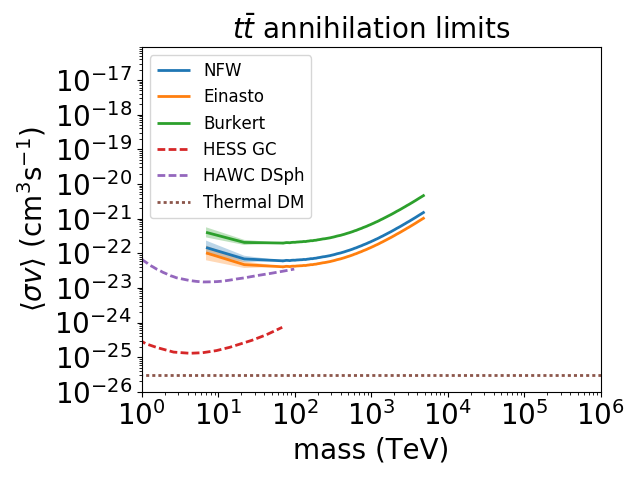}
  }
  \\
 \subfigure{
    \hspace{-0.3cm}\includegraphics[width=7.9cm]{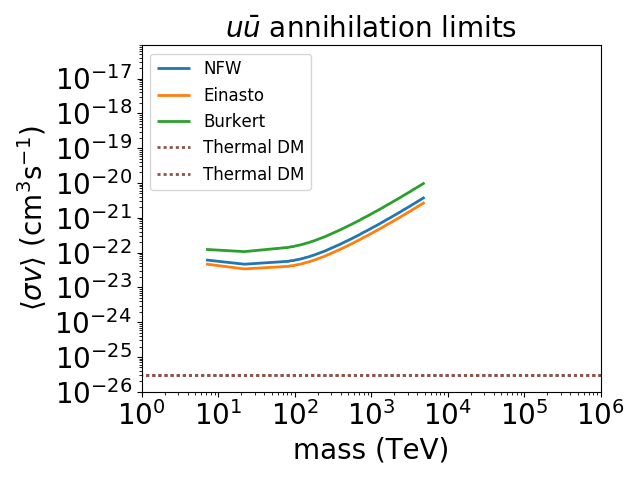}
  }
 \subfigure{
    \hspace{-0.5cm}\includegraphics[width=7.9cm]{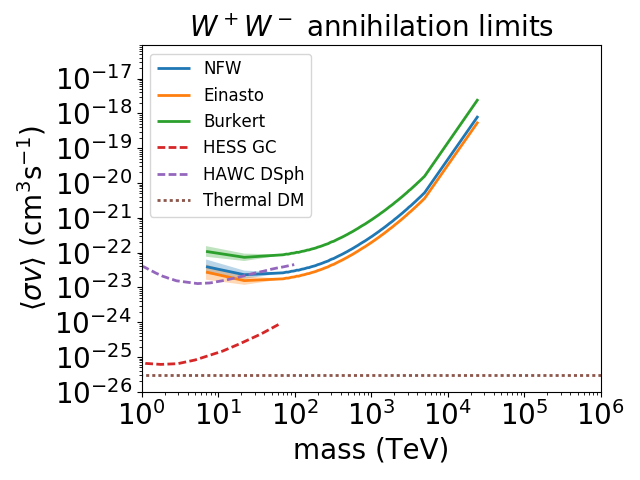}
  }
  \\
  \subfigure{
    \hspace{-0.3cm}\includegraphics[width=7.9cm]{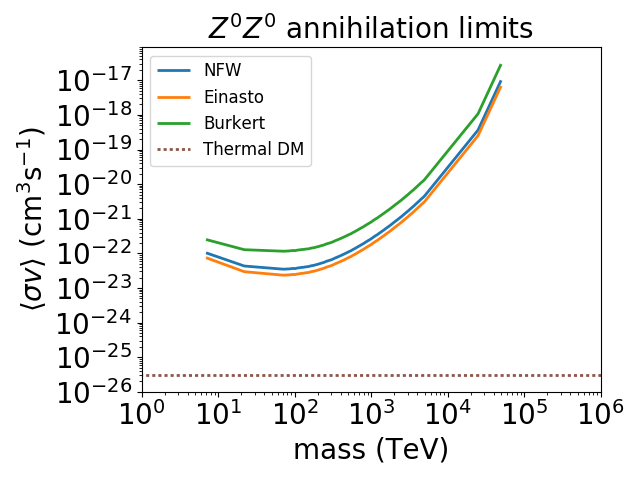}
  }
 \subfigure{
    \hspace{-0.5cm}\includegraphics[width=7.9cm]{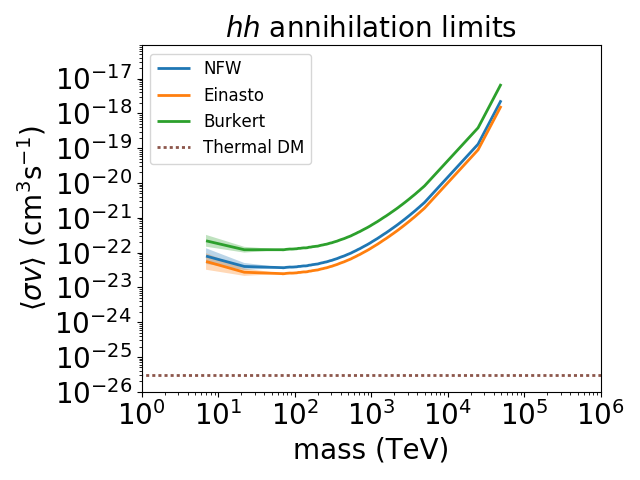}
  }
 \vspace{-0.5cm}
  \caption[DM annihilation limits 1]{95$\%$ confidence level upper limits on the DM annihilation cross-section for the hadronic and bosonic DM annihilation channels considered in this analysis. Limits for a Burkert (upper, green), Einasto (lower, orange), and NFW (middle, blue) profile are shown. GC limits from HESS (for Einasto profile) are shown near the bottom dashed in red. HAWC limits from dwarf galaxies are shown near the middle  dashed in purple. The thermal dark matter cross-section is shown in brown for comparison (bottom).
\label{annlim1}}
\end{figure*}

\begin{figure*}[t]
  \subfigure{
    \hspace{-0.3cm}\includegraphics[width=7.9cm]{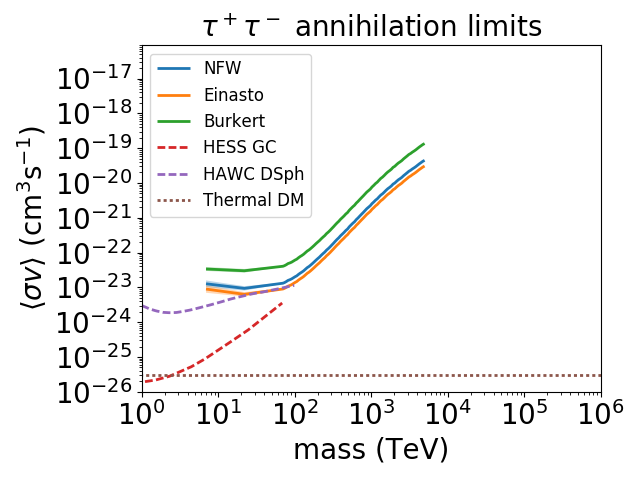}
  }
  \subfigure{
    \hspace{-0.5cm}\includegraphics[width=7.9cm]{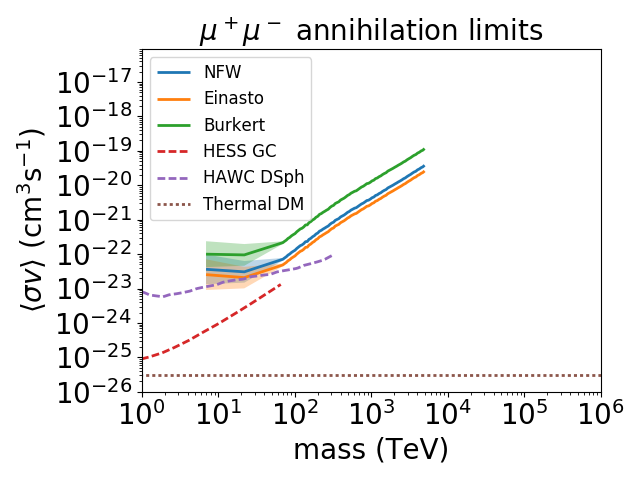}
  }
  \\
 \subfigure{
    \hspace{-0.3cm}\includegraphics[width=7.9cm]{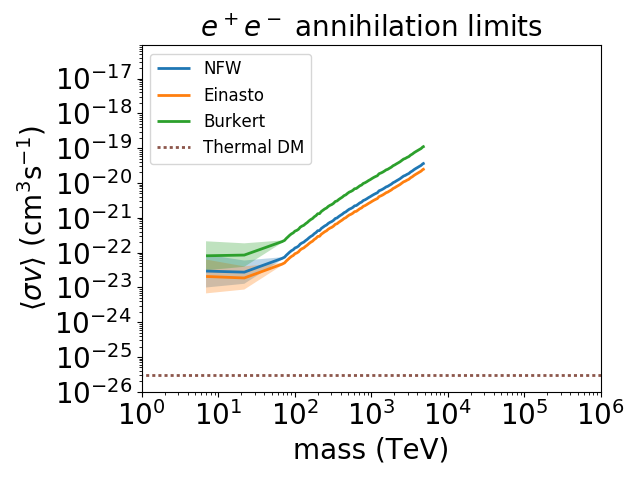}
  }
 \subfigure{
    \hspace{-0.5cm}\includegraphics[width=7.9cm]{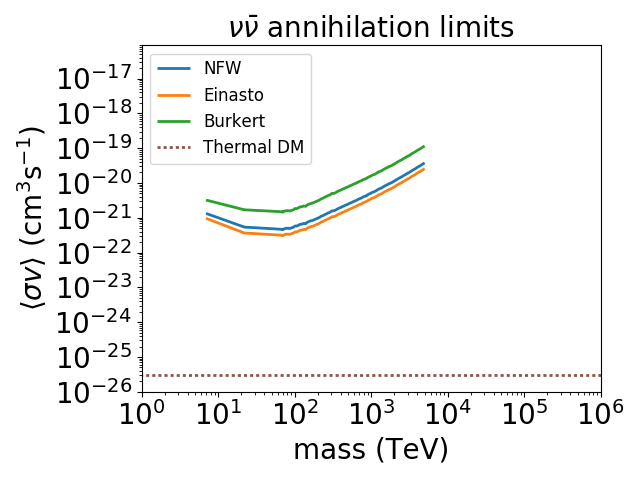}
  }
  \caption[DM annihilation limits 2]{95$\%$ confidence level upper limits on the DM annihilation cross-section for the leptonic and neutrino DM annihilation channels considered in this analysis. Limits for a Burkert (upper, green), Einasto (lower, orange), and NFW (middle, blue) profile are shown. All additional lines are as in Figure~\ref{annlim1}.
\label{annlim2}}
\end{figure*}

\begin{figure*}[t]
  \subfigure{
    \hspace{-0.3cm}\includegraphics[width=7.9cm]{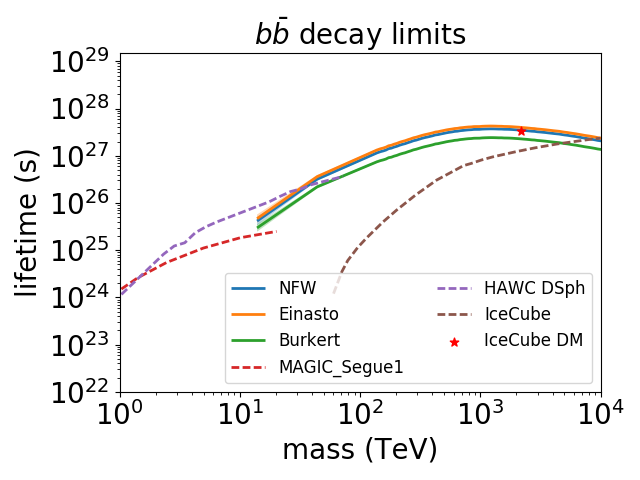}
  }
  \subfigure{
    \hspace{-0.5cm}\includegraphics[width=7.9cm]{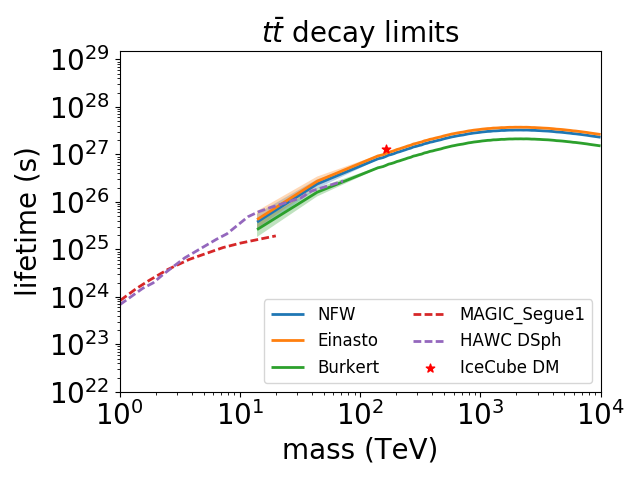}
  }
  \\
 \subfigure{
    \hspace{-0.3cm}\includegraphics[width=7.9cm]{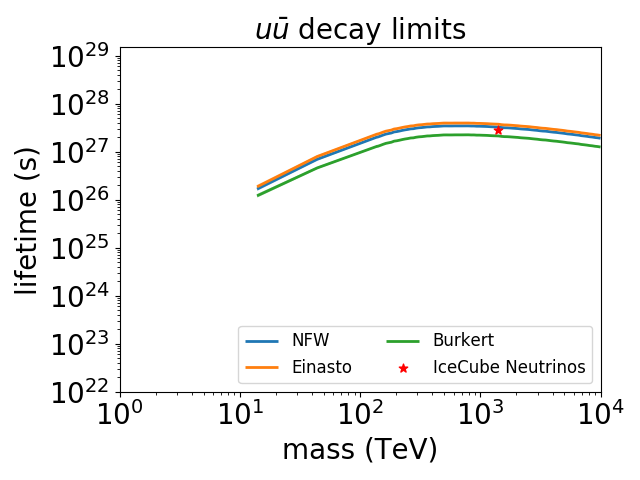}
  }
 \subfigure{
    \hspace{-0.5cm}\includegraphics[width=7.9cm]{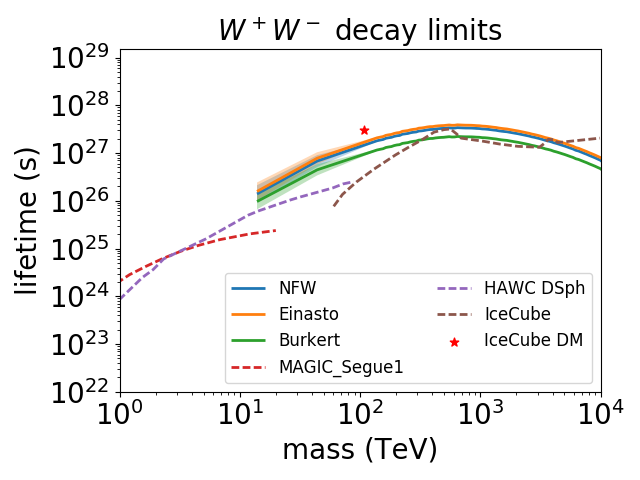}
  }
  \\
  \subfigure{
    \hspace{-0.3cm}\includegraphics[width=7.9cm]{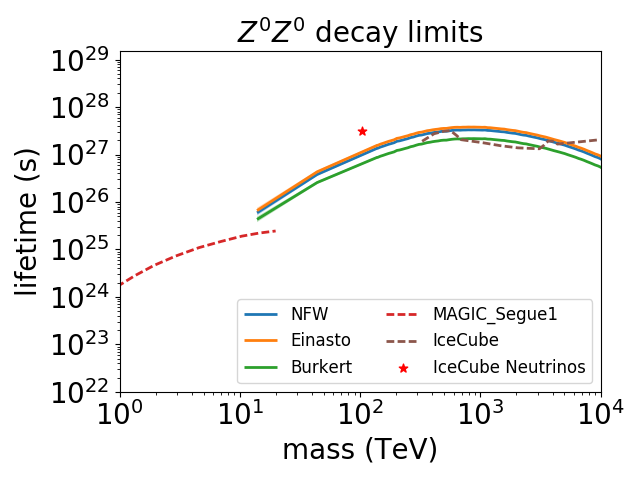}
  }
 \subfigure{
    \hspace{-0.5cm}\includegraphics[width=7.9cm]{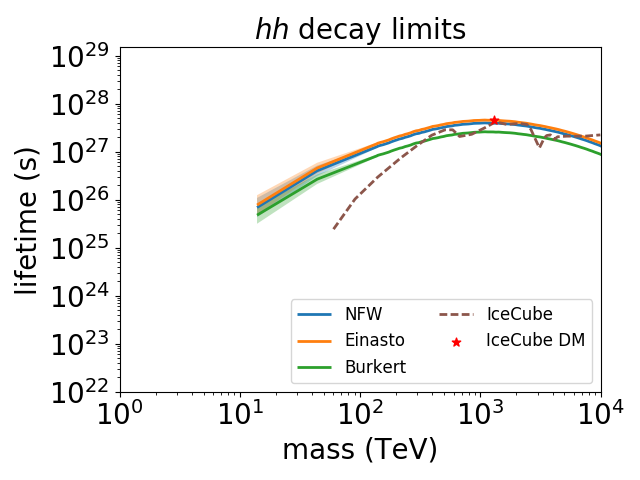}
  }
 \vspace{-0.5cm}
  \caption[DM decay limits 1]{95$\%$ confidence level lower limits on the DM decay lifetime for the hadronic and bosonic DM annihilation channels considered in this analysis. Limits for a Burkert (bottom, green), Einasto (top, orange), and NFW (middle, blue) profile are shown. Segue 1 dwarf galaxy limits from MAGIC~\cite{Aleksic:2013xea} (lower left) are dashed in red. HAWC limits from dwarf galaxies (upper left) are shown dashed in purple. IceCube limits~\cite{Esmaili:2014rma} are shown in brown (on right). The red star is a model in which the IceCube neutrinos are produced in decaying DM~\cite{Cohen:2016uyg}.
\label{declim1}}
\end{figure*}

\begin{figure*}[t]
  \subfigure{
    \hspace{-0.3cm}\includegraphics[width=7.9cm]{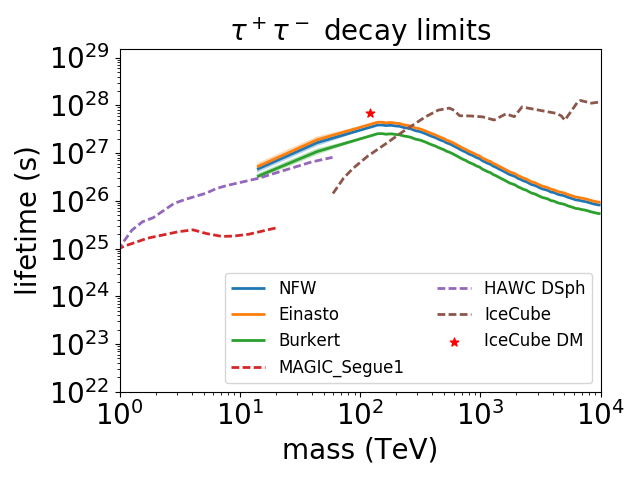}
  }
  \subfigure{
    \hspace{-0.5cm}\includegraphics[width=7.9cm]{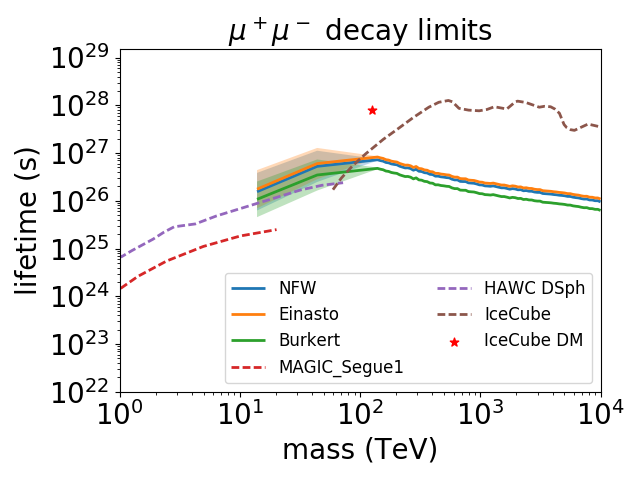}
  }\\
 \subfigure{
    \hspace{-0.3cm}\includegraphics[width=7.9cm]{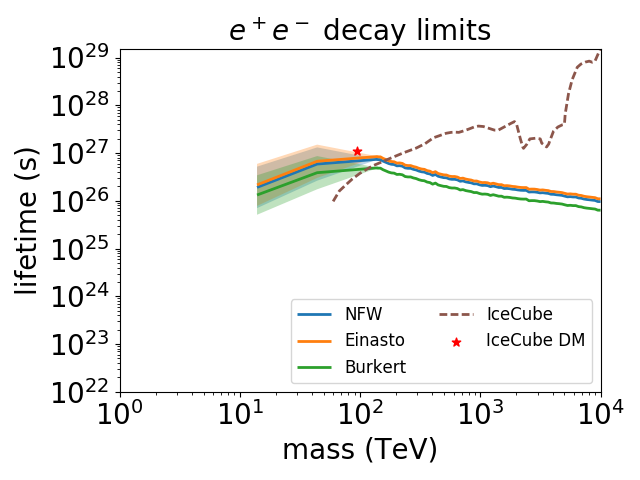}
  }
 \subfigure{
    \hspace{-0.5cm}\includegraphics[width=7.9cm]{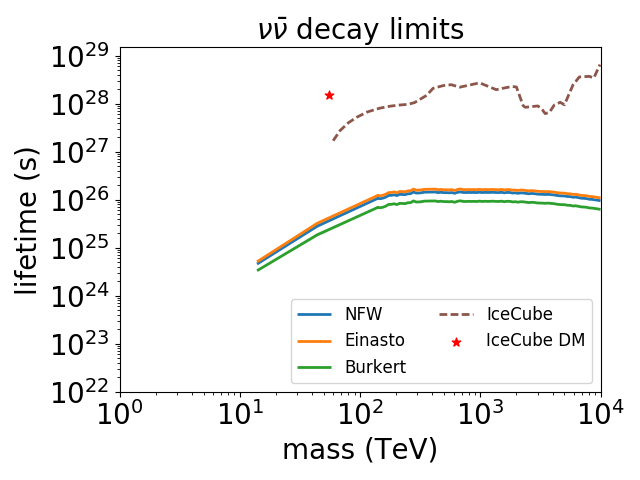}
  }
  \caption[DM decay limits 2]{95$\%$ confidence level lower limits on the DM decay lifetime for the leptonic and neutrino DM annihilation channels considered in this analysis. Limits for a Burkert (bottom, green), Einasto (top, orange), and NFW (middle, blue) profile are shown. All additional lines are as in Figure~\ref{declim1}.
\label{declim2}}
\end{figure*}

Following the methods of the previous section, we calculate limits on the DM annihilation cross-section and decay lifetime for HAWC observations near the GC. We show the limits for ten DM annihilation and decay channels: $u\bar{u}$, $b\bar{b}$, $t\bar{t}$, $W^+W^-$, $ZZ$, $hh$, $e^+e^-$, $\mu^+\mu^-$, $\tau^+\tau^-$, and $\nu\nu$ (an average of the three neutrino species). The annihilation limits are shown in Figures~\ref{annlim1}-\ref{annlim2} and the decay limits are shown in Figures~\ref{declim1}-\ref{declim2}.

\begin{figure*}[t]
  \subfigure{
    \hspace{-0.3cm}\includegraphics[width=7.9cm]{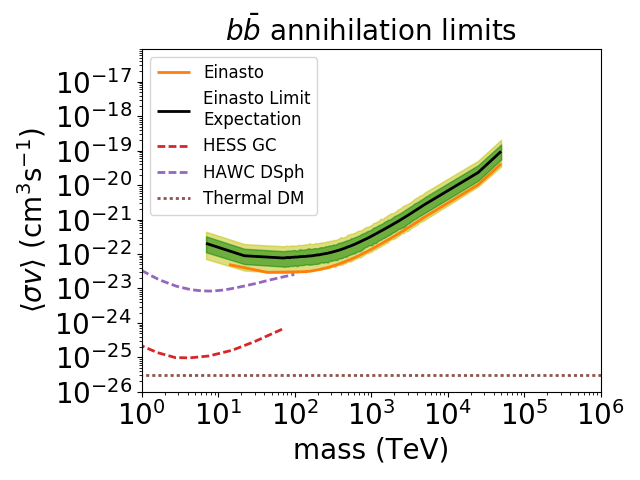}
  }
  \subfigure{
    \hspace{-0.5cm}\includegraphics[width=7.9cm]{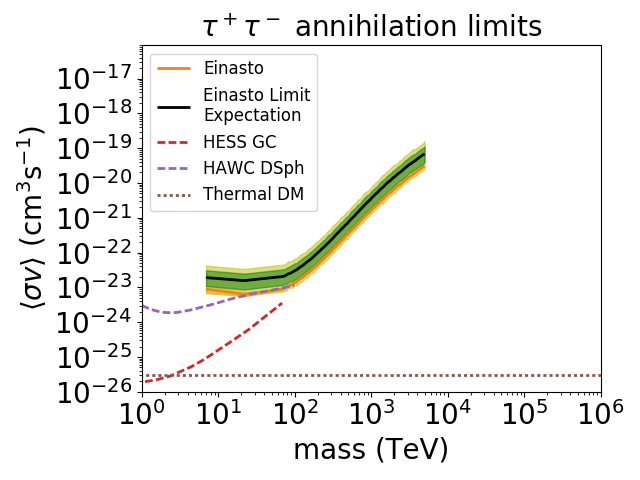}
  }\\
 \subfigure{
    \hspace{-0.3cm}\includegraphics[width=7.9cm]{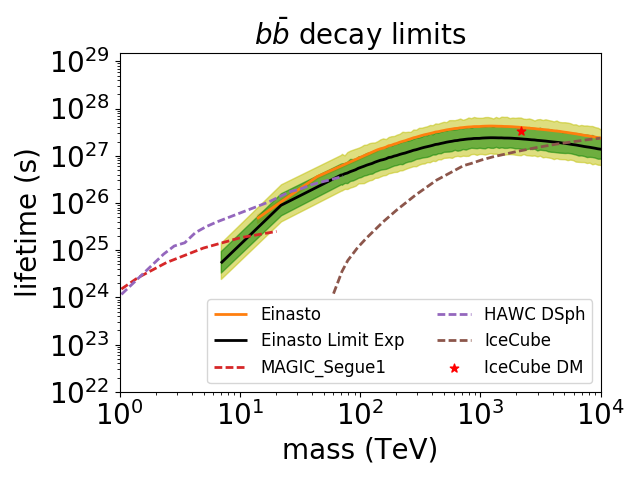}
  }
 \subfigure{
    \hspace{-0.5cm}\includegraphics[width=7.9cm]{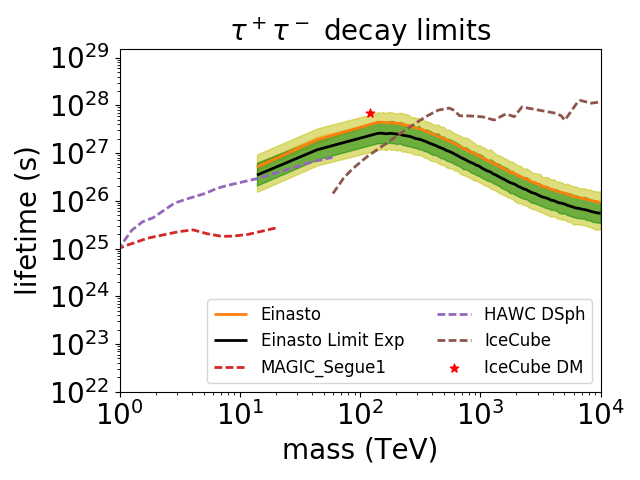}
  }
  \caption[Limit Expectations]{Demonstration of the statistical errors compared to the limits for $b\bar{b}$ and $\tau^+\tau^-$ DM annihilation and decay channels considered in this analysis. Limits for an Einasto (orange) profile are shown as in Figures~\ref{annlim1}-\ref{declim2}. Segue 1 dwarf galaxy limits from MAGIC~\cite{Aleksic:2013xea}, GC limits from HESS~\cite{Abdallah:2016ygi}, HAWC limits from dwarf galaxies, and the model for IceCube neutrinos from decaying DM~\cite{Cohen:2016uyg} are also shown as in Figures~\ref{annlim1}-\ref{declim2}. The black line is the expected value for the Einasto profile without any statistical fluctuations in the data. The green and yellow regions show the limits expected from 68 percent and 95 percent fluctuations from the black line. Note that the values observed in data are underfluctuations lying between the 68 and 95-percent regions.
}\label{sensanndec}
\end{figure*}

The data in the region-of-interest chosen show a slight deficit compared to background, at roughly 1-1.5 sigma. Therefore, the limits shown are slightly better than would be predicted based purely on the HAWC sensitivity. Therefore, we show this predicted limit in Figure~\ref{sensanndec} as a black line. These are shown for the $b\bar{b}$ and $\tau^+\tau^-$ DM profiles. Surrounding the black line is a green band and yellow band. These correspond to the 68- and 95-percent fluctuations away from the black line. This data has under-fluctuations for this region on the sky. In addition to this statistical uncertainty, it should be noted that HAWC has a systematic uncertainty of 50\% at these high energies~\cite{HAWCCrab}.

\section{Discussion}
The limits in Figures~\ref{annlim1}-\ref{annlim2} are shown for three different DM profiles, the peaked NFW and Einasto profiles and the cored Burkert profile. However, because this analysis is not sensitive to the DM flux directly at the GC, these limits change by less than a factor of three between the three DM profiles. In this way, these limits are robust and much less dependent on the DM profile than Galactocentric limits done in the immediate vicinity of the GC, such as the limits from HESS~\cite{Abdallah:2016ygi}.

As discussed in Ref.~\cite{Albert:2017vtb}, the limits obtained using this analysis are approximate, due to the finite energy range of each energy bin. These effects dominate near the edge of each energy bin. Particularly for extremely hard spectra, such as monochromatic gamma-ray lines, the spectral limits can change significantly as a spectral cutoff shifts to energies outside of each energy bin. In order to correctly account for these abrupt unphysical discontinuities, we treat these as systematic errors, which we show as shaded bands around our limits. 

The systematic error bands for the curves are small for softer spectra, including the hadronic and bosonic channels. Even for the fairly hard spectra of $\tau^+\tau^-$ and $\nu\bar{\nu}$ channels, the systematic band is small. However, for the hardest channels, $\mu^+\mu^-$ and $e^+e^-$, the spectrum cuts off sharply, making it fairly monochromatic and resulting in a wider systematic error band. This is a feature of all analyses of monochromatic (or nearly-monochromatic) spectra analyzed using energy bin-based analyses, particularly those with large energy uncertainties. As the energy uncertainty decreases and the energy bins become narrower, this systematic error is reduced.

\subsection{Dark Matter Annihilation Limits}

The DM annihilation limits are shown in Figures~\ref{annlim1}-\ref{annlim2}. Although the GC peaks at a zenith angle of 48 degrees, a large zenith angle for HAWC, the large expected gamma-ray flux from DM from the GC still provides strong limits on the DM annihilation cross-section. While this analysis is only calculated for energies below 39 TeV, the limits are within a factor of 2 of the anticipated HAWC sensitivity plots of Ref.~\cite{Abeysekara:2014ffg} at 1-10 TeV DM masses. This is consistent with those plots being 5-year sensitivity curves and these limits made with less than two years of data. 

For energies higher than tens of TeV, the current HAWC analysis has large systematic errors, so we limit the results here to energies below 39 TeV. More precise energy estimators are being vetted which will significantly improve the understanding of HAWC energies above this value. The precise HAWC sensitivity above these energies is currently under study, but the limits at higher masses should improve by at least a factor of 3 with HAWC flux limits up to 80 TeV. A more optimized analysis of the region-of-interest surrounding the DM signal and more optimized choice of background region should improve these limits further as well.

Below hundreds of TeV in DM mass, these annihilation limits are not as strong as those from the HESS collaboration study of the GC~\cite{Abdallah:2016ygi}. However, our limits are much less sensitive to the DM profile, as discussed above. Therefore, these limits complement current limits for cases where the DM halo is cored, such as a Burkert scenario. Additionally, these conservative limits are within factor of 3 of HESS limits at 100 TeV for leptonic DM channels. For hadronic and bosonic DM channels, our limits are within a factor of 30 of HESS limits at 100 TeV and should be within a factor of three of the HESS limits above 1 PeV, extrapolating the published HESS limits. Therefore, at these high masses, the DM is being well-constrained regardless of DM profile. However, it should be noted that our limits are still only probing cross-sections 2-3 orders of magnitude larger than for thermal DM.

\subsection{Dark Matter Decay Limits}

The DM decay limits are shown in Figures~\ref{declim1}-\ref{declim2}. For DM decay, which has a more spatially-extended profile than annihilation, these HAWC limits are some of the most constraining. More pointed instruments typically cannot observe the wide fields-of-view needed to do these limits with the GC and have instead focused on Galaxy clusters or dwarf galaxies for DM decay limits. Compared to one of the strongest of these, the MAGIC observations of the Segue 1 dwarf galaxy~\cite{Aleksic:2013xea}, the HAWC limits are stronger over the entire mass range studied, above 14 TeV DM mass. 

Limits based on IceCube observations of the Galactic halo also can constrain PeV-level DM masses. In particular, Ref.~\cite{Esmaili:2014rma} showed that neutrino limits are competitive with gamma-ray limits on DM for DM masses above 100 TeV. In Figures~\ref{declim1}-\ref{declim2}, these are shown as the brown lines. Below 100 TeV, the IceCube sensitivity falls quickly, so our limits are much better. However, our limits are similar to the IceCube limits from 1-10 PeV in DM mass for the hadronic and bosonic DM channels. For leptonic and neutrino dark matter channels, many neutrinos are produced, so the IceCube limits are stronger than those in our analysis.

A study of {\it Fermi}-LAT data that includes Inverse Compton emission from charged particles emitted by the DM decays has also shown strong DM decay limits in this mass range~\cite{Cohen:2016uyg}. Most of the limits in this paper are 1-2 orders of magnitude worse than the {\it Fermi}-LAT limit, but our analysis only includes the prompt gamma-ray emission from the decays. Our analysis is insensitive to the details of particle propagation and the low-energy tails of the decay spectra. Our limits can be viewed as more robust, more conservative limits to complement theirs. 
Also, at the highest masses, the addition of higher energy bins, more high-energy data, and better analysis in the future should improve the HAWC limits significantly and make these conservative limits even more constraining.

Also included in these plots is an interpretation of the IceCube neutrino excess coming from DM~\cite{Cohen:2016uyg}. For the hadronic channels, all three DM profiles are in tension with this interpretation of the IceCube excess. For bosonic profiles, the more peaked dark matter profiles do constrain a DM interpretation of the IceCube excess, while the less-peaked Burkert profile does not quite. In all three DM profiles, the decay to $Z$ bosons is currently allowed, though the HAWC limits are within tens of percent of testing that hypothesis. For the leptonic channels, which readily produce neutrinos, the HAWC limits do not yet constrain the IceCube excess. However, the limits on decay into $\tau$ leptons and $e^+e^-$ will test those hypotheses soon. 

\section{Conclusions}

In this analysis, we searched for DM annihilation and decay signals coming from near the GC with HAWC. Due to the lack of significant excess in this region, it has been possible to constrain the DM annihilation cross-section and decay lifetime for TeV-PeV DM masses. These limits were robust across different DM profiles, allowing us to make morphology-independent constraints on the DM. Particularly for DM decay, we provide some of the strongest and most robust limits on the DM decay lifetime. With improved understanding of the HAWC high-energy sensitivity and optimized DM analysis, HAWC DM searches from the GC should be even stronger in the future.

\acknowledgments
We acknowledge the support from: the US National Science Foundation (NSF) the US Department of Energy Office of High-Energy Physics; 
the Laboratory Directed Research and Development (LDRD) program of Los Alamos National Laboratory; 
Consejo Nacional de Ciencia y Tecnolog\'{\i}a (CONACyT), M{\'e}xico (grants 271051, 232656, 260378, 179588, 239762, 254964, 271737, 258865, 243290, 132197, 281653)(C{\'a}tedras 873, 1563), Laboratorio Nacional HAWC de rayos gamma; 
L'OREAL Fellowship for Women in Science 2014; 
Red HAWC, M{\'e}xico; 
DGAPA-UNAM (grants IG100317, IN111315, IN111716-3, IA102715, 109916, IA102917); 
VIEP-BUAP; 
PIFI 2012, 2013, PROFOCIE 2014, 2015; 
the University of Wisconsin Alumni Research Foundation; 
the Institute of Geophysics, Planetary Physics, and Signatures at Los Alamos National Laboratory; 
Polish Science Centre grant DEC-2014/13/B/ST9/945; 
Coordinaci{\'o}n de la Investigaci{\'o}n Cient\'{\i}fica de la Universidad Michoacana. Thanks to Scott Delay, Luciano D\'{\i}az and Eduardo Murrieta for technical support.
BRS is supported by a Pappalardo Fellowship in Physics at MIT. NLR is supported by DOE grants DE-SC00012567 and DE-SC0013999. 

\bibliography{DMTheory,HAWCstuff}

\providecommand{\href}[2]{#2}\begingroup\raggedright\begin{thebibliography}{10}

\bibitem{Faraggi:2000pv}
A.~E. Faraggi and M.~Pospelov, {\it {Selfinteracting dark matter from the
  hidden heterotic string sector}},  {\em Astropart. Phys.} {\bf 16} (2002)
  451--461, [\href{http://arxiv.org/abs/hep-ph/0008223}{{\tt hep-ph/0008223}}].

\bibitem{Boddy:2014yra}
K.~K. Boddy, J.~L. Feng, M.~Kaplinghat, and T.~M.~P. Tait, {\it
  {Self-Interacting Dark Matter from a Non-Abelian Hidden Sector}},  {\em Phys.
  Rev.} {\bf D89} (2014), no.~11 115017,
  [\href{http://arxiv.org/abs/1402.3629}{{\tt arXiv:1402.3629}}].

\bibitem{Forestell:2016qhc}
L.~Forestell, D.~E. Morrissey, and K.~Sigurdson, {\it {Non-Abelian Dark Forces
  and the Relic Densities of Dark Glueballs}},  {\em Phys. Rev.} {\bf D95}
  (2017), no.~1 015032, [\href{http://arxiv.org/abs/1605.08048}{{\tt
  arXiv:1605.08048}}].

\bibitem{Halverson:2016nfq}
J.~Halverson, B.~D. Nelson, and F.~Ruehle, {\it {String Theory and the Dark
  Glueball Problem}},  {\em Phys. Rev.} {\bf D95} (2017), no.~4 043527,
  [\href{http://arxiv.org/abs/1609.02151}{{\tt arXiv:1609.02151}}].

\bibitem{Acharya:2017szw}
B.~S. Acharya, M.~Fairbairn, and E.~Hardy, {\it {Glueball dark matter in
  non-standard cosmologies}},  \href{http://arxiv.org/abs/1704.01804}{{\tt
  arXiv:1704.01804}}.

\bibitem{Soni:2017nlm}
A.~Soni, H.~Xiao, and Y.~Zhang, {\it {A Cosmic Selection Rule for Glueball Dark
  Matter Relic Density}},  \href{http://arxiv.org/abs/1704.02347}{{\tt
  arXiv:1704.02347}}.

\bibitem{Cohen:2016uyg}
T.~Cohen, K.~Murase, N.~L. Rodd, B.~R. Safdi, and Y.~Soreq, {\it {Gamma-ray
  Constraints on Decaying Dark Matter and Implications for IceCube}},
  \href{http://arxiv.org/abs/1612.05638}{{\tt arXiv:1612.05638}}.

\bibitem{Griest:1989wd}
K.~Griest and M.~Kamionkowski, {\it {Unitarity Limits on the Mass and Radius of
  Dark Matter Particles}},  {\em Phys. Rev. Lett.} {\bf 64} (1990) 615.

\bibitem{Hui:2001wy}
L.~Hui, {\it {Unitarity bounds and the cuspy halo problem}},  {\em Phys. Rev.
  Lett.} {\bf 86} (2001) 3467--3470,
  [\href{http://arxiv.org/abs/astro-ph/0102349}{{\tt astro-ph/0102349}}].

\bibitem{Beacom:2006tt}
J.~F. Beacom, N.~F. Bell, and G.~D. Mack, {\it {General Upper Bound on the Dark
  Matter Total Annihilation Cross Section}},  {\em Phys. Rev. Lett.} {\bf 99}
  (2007) 231301, [\href{http://arxiv.org/abs/astro-ph/0608090}{{\tt
  astro-ph/0608090}}].

\bibitem{Berlin:2016gtr}
A.~Berlin, D.~Hooper, and G.~Krnjaic, {\it {Thermal Dark Matter From A Highly
  Decoupled Sector}},  {\em Phys. Rev.} {\bf D94} (2016), no.~9 095019,
  [\href{http://arxiv.org/abs/1609.02555}{{\tt arXiv:1609.02555}}].

\bibitem{Berlin:2016vnh}
A.~Berlin, D.~Hooper, and G.~Krnjaic, {\it {PeV-Scale Dark Matter as a Thermal
  Relic of a Decoupled Sector}},  {\em Phys. Lett.} {\bf B760} (2016) 106--111,
  [\href{http://arxiv.org/abs/1602.08490}{{\tt arXiv:1602.08490}}].

\bibitem{Aartsen:2013bka}
{\bf IceCube} Collaboration, M.~G. Aartsen et~al., {\it {First observation of
  PeV-energy neutrinos with IceCube}},  {\em Phys. Rev. Lett.} {\bf 111} (2013)
  021103, [\href{http://arxiv.org/abs/1304.5356}{{\tt arXiv:1304.5356}}].

\bibitem{Aartsen:2013jdh}
{\bf IceCube} Collaboration, M.~G. Aartsen et~al., {\it {Evidence for
  High-Energy Extraterrestrial Neutrinos at the IceCube Detector}},  {\em
  Science} {\bf 342} (2013) 1242856,
  [\href{http://arxiv.org/abs/1311.5238}{{\tt arXiv:1311.5238}}].

\bibitem{Aartsen:2015knd}
{\bf IceCube} Collaboration, M.~G. Aartsen et~al., {\it {A combined
  maximum-likelihood analysis of the high-energy astrophysical neutrino flux
  measured with IceCube}},  {\em Astrophys. J.} {\bf 809} (2015), no.~1 98,
  [\href{http://arxiv.org/abs/1507.03991}{{\tt arXiv:1507.03991}}].

\bibitem{Aartsen:2015rwa}
{\bf IceCube} Collaboration, M.~G. Aartsen et~al., {\it {Evidence for
  Astrophysical Muon Neutrinos from the Northern Sky with IceCube}},  {\em
  Phys. Rev. Lett.} {\bf 115} (2015), no.~8 081102,
  [\href{http://arxiv.org/abs/1507.04005}{{\tt arXiv:1507.04005}}].

\bibitem{Aartsen:2013uuv}
{\bf IceCube} Collaboration, M.~G. Aartsen et~al., {\it {Search for
  Time-independent Neutrino Emission from Astrophysical Sources with 3 yr of
  IceCube Data}},  {\em Astrophys. J.} {\bf 779} (2013) 132,
  [\href{http://arxiv.org/abs/1307.6669}{{\tt arXiv:1307.6669}}].

\bibitem{Aartsen:2014cva}
{\bf IceCube} Collaboration, M.~G. Aartsen et~al., {\it {Searches for Extended
  and Point-like Neutrino Sources with Four Years of IceCube Data}},  {\em
  Astrophys. J.} {\bf 796} (2014), no.~2 109,
  [\href{http://arxiv.org/abs/1406.6757}{{\tt arXiv:1406.6757}}].

\bibitem{Aartsen:2016oji}
{\bf IceCube} Collaboration, M.~G. Aartsen et~al., {\it {All-sky Search for
  Time-integrated Neutrino Emission from Astrophysical Sources with 7 yr of
  IceCube Data}},  {\em Astrophys. J.} {\bf 835} (2017), no.~2 151,
  [\href{http://arxiv.org/abs/1609.04981}{{\tt arXiv:1609.04981}}].

\bibitem{Hooper:2016jls}
D.~Hooper, {\it {A Case for Radio Galaxies as the Sources of IceCube's
  Astrophysical Neutrino Flux}},  {\em JCAP} {\bf 1609} (2016), no.~09 002,
  [\href{http://arxiv.org/abs/1605.06504}{{\tt arXiv:1605.06504}}].

\bibitem{Murase:2016gly}
K.~Murase and E.~Waxman, {\it {Constraining High-Energy Cosmic Neutrino
  Sources: Implications and Prospects}},  {\em Phys. Rev.} {\bf D94} (2016),
  no.~10 103006, [\href{http://arxiv.org/abs/1607.01601}{{\tt
  arXiv:1607.01601}}].

\bibitem{Esmaili:2013gha}
A.~Esmaili and P.~D. Serpico, {\it {Are IceCube neutrinos unveiling PeV-scale
  decaying dark matter?}},  {\em JCAP} {\bf 1311} (2013) 054,
  [\href{http://arxiv.org/abs/1308.1105}{{\tt arXiv:1308.1105}}].

\bibitem{Feldstein:2013kka}
B.~Feldstein, A.~Kusenko, S.~Matsumoto, and T.~T. Yanagida, {\it {Neutrinos at
  IceCube from Heavy Decaying Dark Matter}},  {\em Phys. Rev.} {\bf D88}
  (2013), no.~1 015004, [\href{http://arxiv.org/abs/1303.7320}{{\tt
  arXiv:1303.7320}}].

\bibitem{Ema:2013nda}
Y.~Ema, R.~Jinno, and T.~Moroi, {\it {Cosmic-Ray Neutrinos from the Decay of
  Long-Lived Particle and the Recent IceCube Result}},  {\em Phys. Lett.} {\bf
  B733} (2014) 120--125, [\href{http://arxiv.org/abs/1312.3501}{{\tt
  arXiv:1312.3501}}].

\bibitem{Zavala:2014dla}
J.~Zavala, {\it {Galactic PeV neutrinos from dark matter annihilation}},  {\em
  Phys. Rev.} {\bf D89} (2014), no.~12 123516,
  [\href{http://arxiv.org/abs/1404.2932}{{\tt arXiv:1404.2932}}].

\bibitem{Bhattacharya:2014vwa}
A.~Bhattacharya, M.~H. Reno, and I.~Sarcevic, {\it {Reconciling neutrino flux
  from heavy dark matter decay and recent events at IceCube}},  {\em JHEP} {\bf
  06} (2014) 110, [\href{http://arxiv.org/abs/1403.1862}{{\tt
  arXiv:1403.1862}}].

\bibitem{Higaki:2014dwa}
T.~Higaki, R.~Kitano, and R.~Sato, {\it {Neutrinoful Universe}},  {\em JHEP}
  {\bf 07} (2014) 044, [\href{http://arxiv.org/abs/1405.0013}{{\tt
  arXiv:1405.0013}}].

\bibitem{Rott:2014kfa}
C.~Rott, K.~Kohri, and S.~C. Park, {\it {Superheavy dark matter and IceCube
  neutrino signals: Bounds on decaying dark matter}},  {\em Phys. Rev.} {\bf
  D92} (2015), no.~2 023529, [\href{http://arxiv.org/abs/1408.4575}{{\tt
  arXiv:1408.4575}}].

\bibitem{Fong:2014bsa}
C.~S. Fong, H.~Minakata, B.~Panes, and R.~Zukanovich~Funchal, {\it {Possible
  Interpretations of IceCube High-Energy Neutrino Events}},  {\em JHEP} {\bf
  02} (2015) 189, [\href{http://arxiv.org/abs/1411.5318}{{\tt
  arXiv:1411.5318}}].

\bibitem{Dudas:2014bca}
E.~Dudas, Y.~Mambrini, and K.~A. Olive, {\it {Monochromatic neutrinos generated
  by dark matter and the seesaw mechanism}},  {\em Phys. Rev.} {\bf D91} (2015)
  075001, [\href{http://arxiv.org/abs/1412.3459}{{\tt arXiv:1412.3459}}].

\bibitem{Ema:2014ufa}
Y.~Ema, R.~Jinno, and T.~Moroi, {\it {Cosmological Implications of High-Energy
  Neutrino Emission from the Decay of Long-Lived Particle}},  {\em JHEP} {\bf
  10} (2014) 150, [\href{http://arxiv.org/abs/1408.1745}{{\tt
  arXiv:1408.1745}}].

\bibitem{Murase:2015gea}
K.~Murase, R.~Laha, S.~Ando, and M.~Ahlers, {\it {Testing the Dark Matter
  Scenario for PeV Neutrinos Observed in IceCube}},  {\em Phys. Rev. Lett.}
  {\bf 115} (2015), no.~7 071301, [\href{http://arxiv.org/abs/1503.04663}{{\tt
  arXiv:1503.04663}}].

\bibitem{Anchordoqui:2015lqa}
L.~A. Anchordoqui, V.~Barger, H.~Goldberg, X.~Huang, D.~Marfatia, L.~H.~M.
  da~Silva, and T.~J. Weiler, {\it {IceCube neutrinos, decaying dark matter,
  and the Hubble constant}},  {\em Phys. Rev.} {\bf D92} (2015), no.~6 061301,
  [\href{http://arxiv.org/abs/1506.08788}{{\tt arXiv:1506.08788}}]. [Erratum:
  Phys. Rev.D94,no.6,069901(2016)].

\bibitem{Boucenna:2015tra}
S.~M. Boucenna, M.~Chianese, G.~Mangano, G.~Miele, S.~Morisi, O.~Pisanti, and
  E.~Vitagliano, {\it {Decaying Leptophilic Dark Matter at IceCube}},  {\em
  JCAP} {\bf 1512} (2015), no.~12 055,
  [\href{http://arxiv.org/abs/1507.01000}{{\tt arXiv:1507.01000}}].

\bibitem{Ko:2015nma}
P.~Ko and Y.~Tang, {\it {IceCube Events from Heavy DM decays through the
  Right-handed Neutrino Portal}},  {\em Phys. Lett.} {\bf B751} (2015) 81--88,
  [\href{http://arxiv.org/abs/1508.02500}{{\tt arXiv:1508.02500}}].

\bibitem{Aisati:2015ova}
C.~El~Aisati, M.~Gustafsson, T.~Hambye, and T.~Scarna, {\it {Dark Matter Decay
  to a Photon and a Neutrino: the Double Monochromatic Smoking Gun Scenario}},
  {\em Phys. Rev.} {\bf D93} (2016), no.~4 043535,
  [\href{http://arxiv.org/abs/1510.05008}{{\tt arXiv:1510.05008}}].

\bibitem{Kistler:2015oae}
M.~D. Kistler, {\it {On TeV Gamma Rays and the Search for Galactic Neutrinos}},
   \href{http://arxiv.org/abs/1511.05199}{{\tt arXiv:1511.05199}}.

\bibitem{Chianese:2016opp}
M.~Chianese, G.~Miele, S.~Morisi, and E.~Vitagliano, {\it {Low energy IceCube
  data and a possible Dark Matter related excess}},  {\em Phys. Lett.} {\bf
  B757} (2016) 251--256, [\href{http://arxiv.org/abs/1601.02934}{{\tt
  arXiv:1601.02934}}].

\bibitem{Fiorentin:2016avj}
M.~Re~Fiorentin, V.~Niro, and N.~Fornengo, {\it {A consistent model for
  leptogenesis, dark matter and the IceCube signal}},  {\em JHEP} {\bf 11}
  (2016) 022, [\href{http://arxiv.org/abs/1606.04445}{{\tt arXiv:1606.04445}}].

\bibitem{Dev:2016qbd}
P.~S.~B. Dev, D.~Kazanas, R.~N. Mohapatra, V.~L. Teplitz, and Y.~Zhang, {\it
  {Heavy right-handed neutrino dark matter and PeV neutrinos at IceCube}},
  {\em JCAP} {\bf 1608} (2016), no.~08 034,
  [\href{http://arxiv.org/abs/1606.04517}{{\tt arXiv:1606.04517}}].

\bibitem{DiBari:2016guw}
P.~Di~Bari, P.~O. Ludl, and S.~Palomares-Ruiz, {\it {Unifying leptogenesis,
  dark matter and high-energy neutrinos with right-handed neutrino mixing via
  Higgs portal}},  {\em JCAP} {\bf 1611} (2016), no.~11 044,
  [\href{http://arxiv.org/abs/1606.06238}{{\tt arXiv:1606.06238}}].

\bibitem{Kalashev:2016cre}
O.~K. Kalashev and M.~{\relax Yu}. Kuznetsov, {\it {Constraining heavy decaying
  dark matter with the high energy gamma-ray limits}},  {\em Phys. Rev.} {\bf
  D94} (2016), no.~6 063535, [\href{http://arxiv.org/abs/1606.07354}{{\tt
  arXiv:1606.07354}}].

\bibitem{Chianese:2016smc}
M.~Chianese and A.~Merle, {\it {A Consistent Theory of Decaying Dark Matter
  Connecting IceCube to the Sesame Street}},  {\em JCAP} {\bf 1704} (2017),
  no.~04 017, [\href{http://arxiv.org/abs/1607.05283}{{\tt arXiv:1607.05283}}].

\bibitem{Aartsen:2014muf}
{\bf IceCube} Collaboration, M.~G. Aartsen et~al., {\it {Atmospheric and
  astrophysical neutrinos above 1 TeV interacting in IceCube}},  {\em Phys.
  Rev.} {\bf D91} (2015), no.~2 022001,
  [\href{http://arxiv.org/abs/1410.1749}{{\tt arXiv:1410.1749}}].

\bibitem{Chianese:2016kpu}
M.~Chianese, G.~Miele, and S.~Morisi, {\it {Dark Matter interpretation of low
  energy IceCube MESE excess}},  {\em JCAP} {\bf 1701} (2017), no.~01 007,
  [\href{http://arxiv.org/abs/1610.04612}{{\tt arXiv:1610.04612}}].

\bibitem{Murase:2015xka}
K.~Murase, D.~Guetta, and M.~Ahlers, {\it {Hidden Cosmic-Ray Accelerators as an
  Origin of TeV-PeV Cosmic Neutrinos}},  {\em Phys. Rev. Lett.} {\bf 116}
  (2016), no.~7 071101, [\href{http://arxiv.org/abs/1509.00805}{{\tt
  arXiv:1509.00805}}].

\bibitem{Palladino:2016xsy}
A.~Palladino, M.~Spurio, and F.~Vissani, {\it {On the IceCube spectral
  anomaly}},  {\em JCAP} {\bf 1612} (2016), no.~12 045,
  [\href{http://arxiv.org/abs/1610.07015}{{\tt arXiv:1610.07015}}].

\bibitem{Abeysekara:2017wzt}
A.~U. Abeysekara et~al., {\it {Search for Very High Energy Gamma Rays from the
  Northern $\textit{Fermi}$ Bubble Region with HAWC}},  {\em Astrophys.J.} {\bf
  842} (2017) 85, [\href{http://arxiv.org/abs/1703.1344}{{\tt
  arXiv:1703.1344}}].

\bibitem{HAWCCrab}
A.~U. Abeysekara et~al., {\it {Observation of the Crab Nebula with the HAWC
  Gamma-Ray Observatory}},  {\em ArXiv} (2017)
  [\href{http://arxiv.org/abs/1701.01778}{{\tt arXiv:1701.01778}}].

\bibitem{Atkins:2003}
R.~Atkins et~al., {\it {Observation of TeV Gamma Rays from the Crab Nebula with
  Milagro Using a New Background Rejection Technique}},  {\em Astrophys.J.}
  {\bf 595} (2003) 803--811, [\href{http://arxiv.org/abs/astro-ph/0305308}{{\tt
  astro-ph/0305308}}].

\bibitem{Abeysekara:2014}
A.~U. Abeysekara et~al., {\it {Observation of Small-scale Anisotropy in the
  Arrival Direction Distribution of TeV Cosmic Rays with HAWC}},  {\em
  Astrophys.J.} {\bf 796} (2014) 108,
  [\href{http://arxiv.org/abs/1408.4805}{{\tt arXiv:1408.4805}}].

\bibitem{Catena:2009mf}
R.~Catena and P.~Ullio, {\it {A novel determination of the local dark matter
  density}},  {\em JCAP} {\bf 1008} (2010) 004,
  [\href{http://arxiv.org/abs/0907.0018}{{\tt arXiv:0907.0018}}].

\bibitem{Navarro:1996gj}
J.~F. Navarro, C.~S. Frenk, and S.~D. White, {\it {A Universal density profile
  from hierarchical clustering}},  {\em Astrophys.J.} {\bf 490} (1997)
  493--508, [\href{http://arxiv.org/abs/astro-ph/9611107}{{\tt
  astro-ph/9611107}}].

\bibitem{Nesti:2013uwa}
F.~Nesti and P.~Salucci, {\it {The Dark Matter halo of the Milky Way, AD
  2013}},  {\em JCAP} {\bf 1307} (2013) 016,
  [\href{http://arxiv.org/abs/1304.5127}{{\tt arXiv:1304.5127}}].

\bibitem{Stadel:2008pn}
J.~Stadel, D.~Potter, B.~Moore, J.~Diemand, P.~Madau, et~al., {\it {Quantifying
  the heart of darkness with GHALO - a multi-billion particle simulation of our
  galactic halo}},  {\em Mon. Not. Roy. Astron. Soc.} {\bf 398} (2009) L21,
  [\href{http://arxiv.org/abs/0808.2981}{{\tt arXiv:0808.2981}}].

\bibitem{Navarro:2008kc}
J.~F. Navarro et~al., {\it {The Diversity and Similarity of Cold Dark Matter
  Halos}},  \href{http://arxiv.org/abs/0810.1522}{{\tt arXiv:0810.1522}}.

\bibitem{Pieri:2009je}
L.~Pieri, J.~Lavalle, G.~Bertone, and E.~Branchini, {\it {Implications of
  High-Resolution Simulations on Indirect Dark Matter Searches}},  {\em Phys.
  Rev.} {\bf D83} (2011) 023518, [\href{http://arxiv.org/abs/0908.0195}{{\tt
  arXiv:0908.0195}}].

\bibitem{Burkert:1995yz}
A.~Burkert, {\it {The Structure of dark matter halos in dwarf galaxies}},  {\em
  IAU Symp.} {\bf 171} (1996) 175,
  [\href{http://arxiv.org/abs/astro-ph/9504041}{{\tt astro-ph/9504041}}].
  [Astrophys. J.447,L25(1995)].

\bibitem{Abeysekara:2017hyn}
A.~U. Abeysekara et~al., {\it {The 2HWC HAWC Observatory Gamma Ray Catalog}},
  {\em Astrophys. J.} {\bf 843} (2017), no.~1 40,
  [\href{http://arxiv.org/abs/1702.02992}{{\tt arXiv:1702.02992}}].

\bibitem{Sjostrand:2006za}
T.~Sjostrand, S.~Mrenna, and P.~Z. Skands, {\it {PYTHIA 6.4 Physics and
  Manual}},  {\em JHEP} {\bf 05} (2006) 026,
  [\href{http://arxiv.org/abs/hep-ph/0603175}{{\tt hep-ph/0603175}}].

\bibitem{Sjostrand:2007gs}
T.~Sjostrand, S.~Mrenna, and P.~Z. Skands, {\it {A Brief Introduction to PYTHIA
  8.1}},  {\em Comput. Phys. Commun.} {\bf 178} (2008) 852--867,
  [\href{http://arxiv.org/abs/0710.3820}{{\tt arXiv:0710.3820}}].

\bibitem{Sjostrand:2014zea}
T.~Sjostrand, S.~Ask, J.~R. Christiansen, R.~Corke, N.~Desai, P.~Ilten,
  S.~Mrenna, S.~Prestel, C.~O. Rasmussen, and P.~Z. Skands, {\it {An
  Introduction to PYTHIA 8.2}},  {\em Comput. Phys. Commun.} {\bf 191} (2015)
  159--177, [\href{http://arxiv.org/abs/1410.3012}{{\tt arXiv:1410.3012}}].

\bibitem{Christiansen:2014kba}
J.~R. Christiansen and T.~Sjostrand, {\it {Weak Gauge Boson Radiation in Parton
  Showers}},  {\em JHEP} {\bf 04} (2014) 115,
  [\href{http://arxiv.org/abs/1401.5238}{{\tt arXiv:1401.5238}}].

\bibitem{Esmaili:2015xpa}
A.~Esmaili and P.~D. Serpico, {\it {Gamma-ray bounds from EAS detectors and
  heavy decaying dark matter constraints}},  {\em JCAP} {\bf 1510} (2015),
  no.~10 014, [\href{http://arxiv.org/abs/1505.06486}{{\tt arXiv:1505.06486}}].

\bibitem{Albert:2017vtb}
A.~Albert et~al., {\it {Dark Matter Limits From Dwarf Spheroidal Galaxies with
  The HAWC Gamma-Ray Observatory}},
  \href{http://arxiv.org/abs/1706.01277}{{\tt arXiv:1706.01277}}.

\bibitem{Aleksic:2013xea}
J.~Aleksić et~al., {\it {Optimized dark matter searches in deep observations
  of Segue 1 with MAGIC}},  {\em JCAP} {\bf 1402} (2014) 008,
  [\href{http://arxiv.org/abs/1312.1535}{{\tt arXiv:1312.1535}}].

\bibitem{Esmaili:2014rma}
A.~Esmaili, S.~K. Kang, and P.~D. Serpico, {\it {IceCube events and decaying
  dark matter: hints and constraints}},  {\em JCAP} {\bf 1412} (2014), no.~12
  054, [\href{http://arxiv.org/abs/1410.5979}{{\tt arXiv:1410.5979}}].

\bibitem{Abdallah:2016ygi}
{\bf H.E.S.S.} Collaboration, H.~Abdallah et~al., {\it {Search for dark matter
  annihilations towards the inner Galactic halo from 10 years of observations
  with H.E.S.S}},  {\em Phys. Rev. Lett.} {\bf 117} (2016), no.~11 111301,
  [\href{http://arxiv.org/abs/1607.08142}{{\tt arXiv:1607.08142}}].

\bibitem{Abeysekara:2014ffg}
{\bf HAWC} Collaboration, A.~U. Abeysekara et~al., {\it {Sensitivity of HAWC to
  high-mass dark matter annihilations}},  {\em Phys. Rev.} {\bf D90} (2014),
  no.~12 122002, [\href{http://arxiv.org/abs/1405.1730}{{\tt
  arXiv:1405.1730}}].

\end{thebibliography}\endgroup

\end{document}